\documentclass[a4paper,11pt]{article} 
\usepackage[margin=25mm]{geometry}
\usepackage{amsmath,graphicx}
\usepackage{amsthm,amssymb}
\usepackage{enumitem}
\usepackage{mwe} 
\usepackage[algo2e,ruled]{algorithm2e}
\usepackage{algorithm}
\usepackage{algpseudocode}
\SetKwInput{KwInput}{Input} 
\SetKwInput{KwOutput}{Output}
\usepackage{graphicx}
\usepackage{blindtext, color,xcolor}
\usepackage{hyperref}
\usepackage{epstopdf,url}
\usepackage{amssymb,nccmath}
\usepackage{stmaryrd}
\usepackage{tabularx}
\usepackage{changepage}
\usepackage{enumitem}
\usepackage{setspace}
\usepackage{multirow}
\usepackage{ragged2e,float}
\usepackage{mathtools,xparse}
\usepackage{relsize}
\usepackage[normalem]{ulem}
\usepackage{setspace}
\usepackage{stmaryrd}
\usepackage{array}
\usepackage[]{mdframed}
\newcolumntype{Y}{>{\centering\arraybackslash}X}
\newcommand{\norm}[1]{\left\lVert#1\right\rVert}
\useunder{\uline}{\ul}{}

\newcommand{\RCH}[1]{{\color{black} #1}}
\newcommand{\RCHH}[1]{{\color{black} #1}}

\providecommand{\keywords}[1]
{
	\small	
	\textbf{{Keywords---}} #1
}

\title{\textbf{Bayesian Multifractal Image Segmentation}}

\author{Kareth~M.~Le\'on-L\'opez$^{1}$, Abderrahim Halimi$^{2}$, \\ Jean-Yves Tourneret$^{1}$, and Herwig Wendt$^{1}$\footnote{This work was partly supported by Grant ANR-18-CE45-0007 MUTATION and the UK Royal Academy of Engineering under the Research Fellowship Scheme (RF/201718/17128).}%
	\\
	{\footnotesize $^{1}$ IRIT Laboratory, Universit\'e de Toulouse, CNRS, INP-Toulouse, UT3, UT2, T\'eSA, Toulouse, France.} 
	\\
	{\footnotesize \texttt{\{firstname.lastname\}@irit.fr}} 
	\\
	{\footnotesize $^{2}$ Heriot-Watt University, Edinburgh EH14 4AS, UK. } 
	\\
	{\footnotesize \texttt{a.halimi@hw.ac.uk}}
}
\date{ } 

\begin{document}
\maketitle  
\begin{abstract}

Multifractal analysis (MFA) provides a framework for the global characterization of image textures by describing the spatial fluctuations of their local regularity based on the multifractal spectrum. Several works have shown the interest of using MFA for the description of homogeneous textures in images. Nevertheless, natural images can be composed of several textures and, in turn, multifractal properties associated with those textures. This paper introduces an unsupervised Bayesian multifractal segmentation method to model and segment multifractal textures by jointly estimating the multifractal parameters and labels on images, at the pixel-level.
For this, a computationally and statistically efficient multifractal parameter estimation model for wavelet leaders is firstly developed, defining different multifractality parameters for different regions of an image. Then, a multiscale Potts Markov random field is introduced as a prior to model the inherent spatial and scale correlations (referred to as cross-scale correlations) between the labels of the wavelet leaders. A Gibbs sampling methodology is finally used to draw samples from the posterior distribution of the unknown model parameters. Numerical experiments are conducted on synthetic multifractal images to evaluate the performance of the proposed segmentation approach. The proposed method achieves superior performance compared to traditional unsupervised segmentation techniques as well as modern deep learning-based approaches, showing its effectiveness for multifractal image segmentation.

\end{abstract}


\section{Introduction}

Scaling, or scale invariance, is a modeling paradigm in which a given object structure or statistical property remains unchanged regardless of the scale at which it is analyzed \cite{abry2013van}, and can be exploited using self-similar or multifractal models \cite{wendt2007bootstrap}. The scaling properties of texture images have been studied via the multifractal analysis (MFA) that allows complex patterns (local transient regularity fluctuations) to be described and modeled across different scales \cite{wendt2007multifractality,wendt2018multifractal}. 
MFA has been exploited in several applications of different natures, including network traffic analysis \cite{abry1998wavelet}, texture classification \cite{wendt2009wavelet,Wendt20091100,Xu2010}, texture characterization \cite{schmitt2001multifractal} and art analysis \cite{abry2013van}. Current state-of-the-art MFA tools rely on the use of dedicated multiresolution coefficients, the wavelet leaders \cite{wendt2007bootstrap,Wendt20091100,jaffard2004wavelet}.

On the other hand, texture segmentation is a long-standing problem in image processing that has received significant research efforts in the past and continues to attract considerable attention in the scientific community \cite{jain1991unsupervised,kirillov2023segment, he2017mask}. Several methods have relied on geometrical or statistical features using Markov random fields \cite{manjunath1991unsupervised,halimi2015unsupervised}, while others rely on pixel intensities to group different regions \cite{yuan2015factorization,haralick1979statistical}. Other works have exploited the scale invariance property for image segmentation \cite{pustelnik2016combining,abadi2006texture}. As an example, self-similarity exponents and local regularity have been used to segment image textures \cite{pustelnik2016combining}. Works such as \cite{chaudhuri1995texture} have also used fractal information to extract features.
\RCH{The literature reports some MFA-based image segmentation methods that rely on the multifractal spectrum \cite{abadi2006texture,san2017application, zhu2019ship} (see definition in Section II-A and see a review of the methods in \cite{lopes2009fractal}). For instance, the multifractal spectrum {was used in \cite{san2017application}} to segment water bodies in synthetic aperture radar (SAR) images. However, {the strategy investigated in \cite{san2017application}} is computationally demanding and relies on an empirical threshold for region delineation. Segmentation approaches based on multifractal descriptors have also been {studied} \cite{xia2006morphology,vehel1994multifractal}. The state-of-the-art MFA formalism (in terms of precision and computational efficiency) is based on wavelet leaders \cite{jaffard2006wavelet,wendt2018multifractal, combrexelle2015bayesian}.}
In this line, authors in \cite{wendt2018multifractal} provide a methodology for modeling the multifractality of multivariate images, yielding an operational tool for the multifractal analysis of multidimensional images as well as images with several multifractal textures. However, the method is not intended to assess multifractality at the pixel level or for segmentation purposes, which is a key limitation of current multifractal-based approaches. Indeed, characterizing point-wise regularity fluctuations requires to analyze several scales, tied to space averages \cite{wendt2022multifractal}, thus yielding a paradox and an important difficulty for pixel-level analysis and segmentation. 

Multi-resolution coefficients associated with an image have been used in several works to perform image segmentation. {The wavelet transform was used in} \cite{choi2001multiscale} and \cite{brault2005unsupervised} to compute multi-resolution coefficients of an image that are combined from coarse to fine scales for image segmentation. Further works such as \cite{bouman1994multiscale,fan2001joint} have proposed multiscale Bayesian techniques in which a contextual behavior from a coarse scale (low-resolution scale) is employed up to the finest scale (highest resolution) to provide an image segmentation. Multiscale segmentation has also been performed using hidden Markov trees \cite{choi2001multiscale,cheng2001multiscale} or neural networks \cite{kim2009multiscale}. The main limitation of these approaches is that they rely on training data to describe the texture parameters of the images. Additionally, even though there exist works in the literature that attempt to consider coefficients at multiple scales for image segmentation, up to our knowledge, there is no method in the state of the art that exploits these scales to segment images at the pixel level based on a multifractal model.

This work introduces a new multifractal image segmentation approach for the joint parameter and label estimation of heterogeneous images via a multiscale decomposition and a Bayesian framework. 
The proposed segmentation approach is fully unsupervised and is conducted in the wavelet leader domain, in which the scale-free textures are parameterized by the {multifractality} parameter. The contributions of this paper are the following:
\begin{itemize}[label={\footnotesize $\bullet$},leftmargin=5mm]
    \item First, a computationally efficient Fourier-based Whittle approximation for the multifractality parameter estimation on heterogeneous multifractal images is introduced. This contribution relies on the efficient model developed in \cite{wendt2018multifractal} that {is extended and adapted} to \textit{irregular \RCH{shapes}} for modeling the different piecewise textures of an image. 
    
    \item Second, a likelihood function defined in terms of the \textit{multifractality parameter} across scales is introduced, leading to a model for scale invariant textures.

    \item Third, a multiscale Potts Markov random field that considers the \textit{intra-}scale and \textit{inter-}scale dependencies in the wavelet leader domain is {considered}. This random field uses a multiscale graph structure that exploits spatial and scale image correlations. Moreover, the granularity coefficients of the proposed multiscale Potts-Markov random field are efficiently and jointly sampled with the labels.

\item Fourth, we conduct extensive numerical experiments and comparisons with state-of-the-art image segmentation methods for simulated and real-world images, showing the interest of the proposed framework. 
\end{itemize}

\section{Multifractal Analysis of Images}

\subsection{Local Regularity and Multifractal Spectrum}

{MFA} enables the study of local regularity of functions. To that end, it makes use of a quantity that measures the so-called H\"older exponent $h(\textbf{t})$ at a location $\textbf{t}$, and a \textit{global} description of its spatial fluctuations, the multifractal spectrum $\mathcal{D}(h)$ \cite{pesquet2002stochastic}.

{\it H\"older Exponent.\quad} 
Let $\textbf{F}(\textbf{t}) : \mathbb{R}^2 \rightarrow \mathbb{R}$ denote the function (image) to be analyzed, under the assumption that $\textbf{F}$ is locally bounded. Then, $\textbf{F}$ is said to belong to $C^{\alpha}(\textbf{t}_0)$ with $\alpha > 0$ if there exists a positive constant $C$ and a polynomial $P_{\textbf{t}_0}$, of degree smaller than $\alpha$, such that $||\textbf{F}(\textbf{t})-P_{\textbf{t}_0}(\textbf{t}) ||\leq C ||\textbf{t}-\textbf{t}_0||^\alpha$ holds, where $\| \cdot \|$ is the Euclidean norm. The H\"older exponent at $\textbf{t}_0$ is then defined as 
$$
    h(\textbf{t}_0) \triangleq \textrm{sup} \{ \alpha : \textbf{F} \in C^{\alpha}(\textbf{t}_0) \} .
$$
Here, large (small) values of $h(\textbf{t}_0) $ correspond to locations where $\mathbf{F}$ is locally smooth (irregular), respectively \cite{pustelnik2016combining}.

{\it Multifractal Spectrum.\quad} The collection of fractal dimensions (Hausdorff dimension $\textrm{dim}_H$) of the sets of points for which the H\"older  exponent takes the same value defines the multifractal spectrum 
\begin{equation}
    \mathcal{D}(h) \triangleq \textrm{dim}_H \{ \textbf{t} : h (\textbf{t}) = h\}  .  \label{eq:spectrum}
\end{equation}
The estimation of $\mathcal{D}(h)$ is the main target of MFA. However, it is unfeasible using its formal definition \eqref{eq:spectrum}. Indeed, the estimation relies on a multiscale decomposition and the so-called \textit{multifractal formalism} (for more details, see \cite{wendt2007bootstrap,wendt2018multifractal,jaffard2004wavelet}). The current state-of-the-art multifractal formalism is based on the wavelet leaders of the two-dimensional (2D) discrete wavelet transform (DWT) coefficients of the given image and is recalled next.

\subsection{Multifractal Formalism based on Wavelet Leaders }

{\it Wavelet Leaders.\quad} \textcolor{black}{The 2D orthonormal DWT of $\textbf{F}\in \mathbb{R}^{N\times N}$ is obtained by considering four 2D filters $\phi^{(m)}(\textbf{n})$, $m=0,...,3$, defined as the tensor products of a low-pass filter $\phi(n)$ (scaling function) and a high-pass filter $\psi(n)$ (wavelet function characterized by $N_{\psi} > 0$ vanishing moments). Precisely, the four 2D filters are written as \cite{combrexelle2015bayesian,wendt2018multifractal}:
$$ 
    \Bigg\{ 
    \begin{matrix}
        \psi^{(0)}(\textbf{n}) = \phi(n_1) \phi(n_2), & \psi^{(1)}(\textbf{n}) = \psi(n_1) \phi(n_2), \\
        \psi^{(2)}(\textbf{n}) = \phi(n_1) \psi(n_2), & \psi^{(3)}(\textbf{n}) = \psi(n_1) \psi(n_2) ,
    \end{matrix}
$$ 
where $\psi^{(0)}(\textbf{n})$ is the 2D low-pass filter and $\{ \psi^{(m)}(\textbf{n}) \}_{m=1}^{3}$ are the 2D high-pass filters. Let $D^{(m)}_{\textbf{F}}(j,\mathbf{n})$ denotes the 2D DWT of $\textbf{F}$, with $D^{(0)}_{\textbf{F}}(0,\mathbf{n}) \triangleq \textbf{F}$. The DWT coefficients $\big\{D_{\textbf{F}}^{(m)}(j,\mathbf{n})\big\}_{m=1}^{3}$, and the approximation coefficients $D_{\textbf{F}}^{(0)}(j,\mathbf{n})$, for $j\geq 1$, are obtained by iteratively convolving $\psi^{(m)}$ with the coefficients $D_{\textbf{F}}^{(0)}(j-1,\cdot)$, followed by decimation \cite{wendt2018multifractal}.
For MFA purposes, the 2D DWT coefficients are normalized as $d_{\textbf{F}}^{(m)}(j,\mathbf{n}) \triangleq 2^{-j} D_{\textbf{F}}^{(m)}(j,\mathbf{n})$, for $m=1,2,3$.}

The \textit{wavelet leaders} are defined \RCH{in \cite{jaffard2006wavelet}} as the largest DWT coefficient magnitudes within the spatial neighborhood $3\lambda_{j,\textbf{n}}$ over all finer scales:
\begin{equation}
    L_{{j,\textbf{n}}} = \underset{
    {\scriptscriptstyle \begin{matrix}
    m \in \{1,2,3\} \\ 
    \lambda' {\subset 3 \lambda_{j,\textbf{n}}}
    \end{matrix} }
    }{\textrm{sup}} \big| {d}^{(m)}_{\lambda'}  \big|, 
    \label{eq_wavleader}
\end{equation}
where $\lambda_{j,\textbf{n}}$ is the dyadic cube of side length $2^j$ at position $2^j \textbf{n}$, i.e., $\lambda_{j,\textbf{n}} = \{[2^j n_1, 2^j (n_1 + 1) ), [2^j n_2, 2^j (n_2 + 1) ) \}$ \cite{lerner2019intuitive,Jaffard2015}, and $3 \lambda_{j,\textbf{n}} = \bigcup_{i_1, i_2 \in \{-1,0,1 \}} \lambda_{j,(n_1 + i_1, n_2 + i_2)} $ \RCH{(see \cite{jaffard2006wavelet} for more details about wavelet leaders and related bounds)}.

\begin{figure}[t]
    \centering    
    \includegraphics[width=0.7\linewidth]{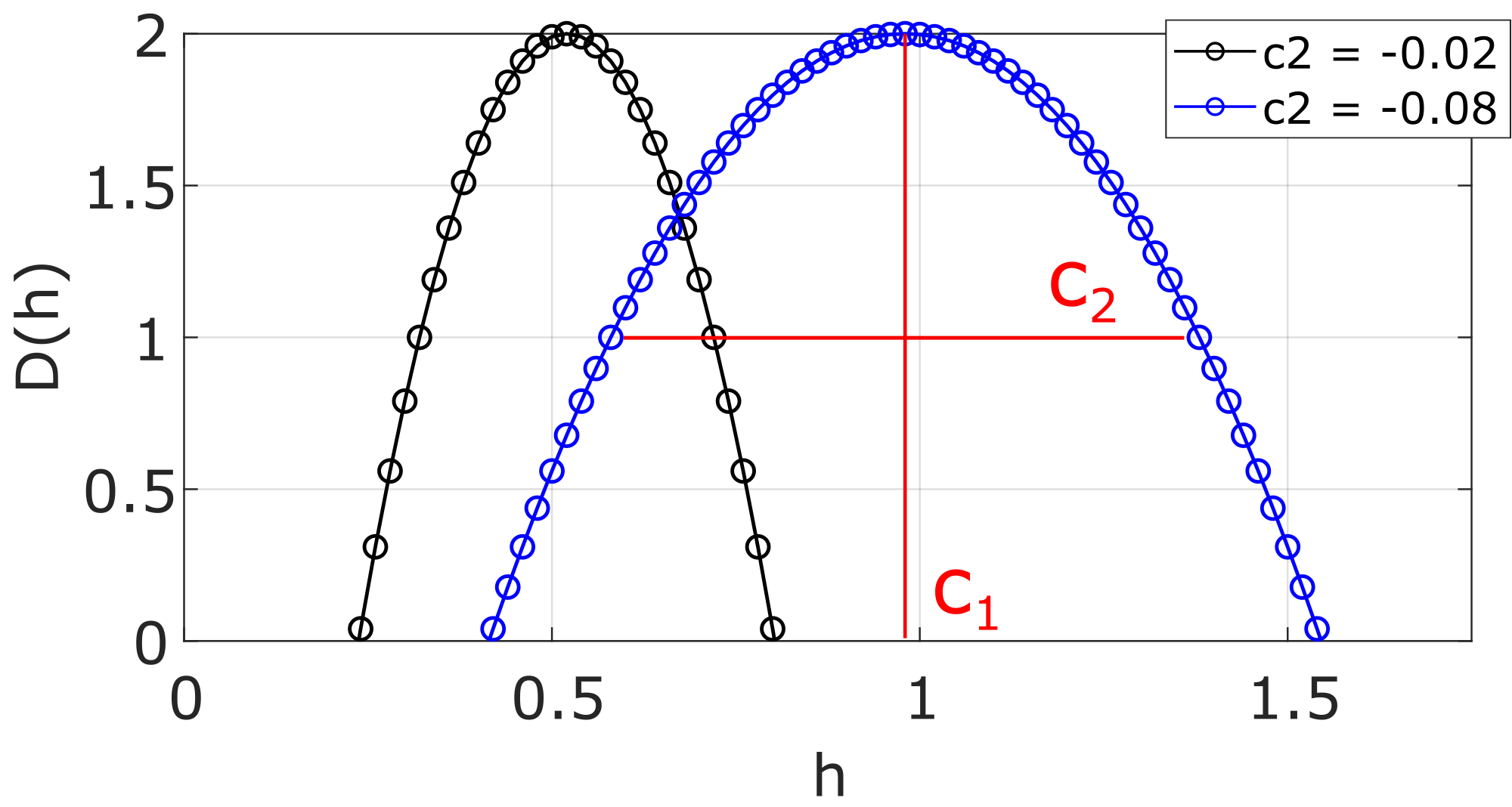}\vspace{-0.3cm}
    \caption{Multifractal spectrum $\mathcal{D}(h)$ schematic representation from two 2D multifractal random walk images, where $c_1$ is the mode and $c_2$, the width, is the multifractality parameter; $(c_1,c_2)$ capture most of the local regularity information of the image.}
    \label{fig:mfs}
\end{figure}

It has been shown that the coefficients $c_p$ are tied to the wavelet leaders $L_{j,\mathbf{n}}$ via the key relation \cite{combrexelle2015bayesian}
\begin{equation}
    \textrm{Cum}_{p}[\ln L_{j,\mathbf{n}} ] = c_p^0 + c_p \ln 2^j.
    \label{eq_cumulant}
\end{equation}
In addition, the multifractal formalism  provides an expansion of $\mathcal{D}(h)$ in terms of the log-cumulants $c_p$, $p\geq 1$, such that
\begin{equation}
    \mathcal{D}(h) = 2+ \frac{c_2}{2!}\bigg(\frac{h-c_1}{c_2}\bigg)^2+ \frac{-c_3}{3!}\bigg(\frac{h-c_2}{c_2}\bigg)^3 + ...   ,
\end{equation}
for $c_2<0$, see \cite{wendt2018multifractal,combrexelle2015bayesian} for further details. 
These so-called log-cumulants provide a relevant summary of $\mathcal{D}(h) $, e.g., $c_1$ is the mode of $\mathcal{D}(h)$ that can be interpreted as the average smoothness of the image $\mathbf{F}$ and is related to the Hurst parameter \cite{mandelbrot1968fractional}, $c_2$ is related to the width of $\mathcal{D}(h)$ and captures the degree of fluctuations of the local regularity of $\mathbf{F}$ and is the most important feature of multifractal functions, termed \emph{multifractality parameter}. \RCH{Figure \ref{fig:mfs} presents a schematic illustration of the theoretical multifractal spectrum for two homogeneous 2D multifractal random walks, shown for $c_1= 0.5, c_2=-0.02$ (blue curve) and $c_1 =0.98 , c_2=-0.08$ (black curve) \cite{wendt2007bootstrap,wendt2007multifractality,Wendt20091100} (note that these are the multifractal spectra of the composed image of Fig. \ref{fig:result_MRW_all_methods}-a-\textit{top}).}

{\it Estimation of $c_2$.\quad} 
The relation \eqref{eq_cumulant} with $p=2$ leads to $ \textrm{Var}[\ln L_{{j,\textbf{n}}}] = c_2^0 + c_2 \ln 2^j $, which indicates that $c_2$ can be estimated by linear regression over a range of scale $j\in [j_1,j_2]$:
\begin{equation}
\label{eq_linear_fit_c2}
    \hat{c}_{2} = \frac{1}{\ln 2} \sum_{j=j_1}^{j_2} w_j \widehat{\textrm{Var}}[\ln (L_{{j,\textbf{n}}})] ,
\end{equation}
where $\widehat{\textrm{Var}}(\cdot)$ denotes the sample variance and $w_j$ are suitable regression weights \cite{wendt2007bootstrap,wendt2007multifractality,abry2000wavelets}. 
The estimator \eqref{eq_linear_fit_c2} has low complexity but poor performance (large variance). To overcome this limitation, a Fourier-based Bayesian model was introduced for homogeneous images  \cite{wendt2018multifractal}. It is well suited for the design of joint priors and briefly recalled next.

\subsection{Estimation of $c_2$ using a Fourier-based Bayesian model} 
\label{subsection:Fourier-based_likelihood}

Let ${\ell}_{j,\textbf{n}} = \ln (L_{{j,\textbf{n}}}) $ denote the centered logarithms of the wavelet leader coefficients, referred to as \textit{log-leaders}, at scale $j$ and at position $\textbf{n}$. The vector of log-leaders, $\boldsymbol{\ell}_{{j}} \in \mathbb{R}^{N_j ^2}$, is obtained by vectorization using the lexicographic order, where $N_{j} = N/2^j$. Note that $j\in \llbracket j_1,j_2 \rrbracket $, $1 \leq j_1 \leq ... \leq j_2 \leq J$, in which $j=1$ and $j=J$ index the finest and the coarsest scales, respectively, where $\llbracket a,b \rrbracket$ denotes the set of integers ranging from $a$ to $b$. Further, let $\boldsymbol{\ell} \triangleq [\boldsymbol{\ell}_{j_1}^T,...,\boldsymbol{\ell}_{j_2}^T]^T$ be the vector obtained by concatenation of the log-leaders at scales $j_1,...,j_2$.
According to \cite{wendt2018multifractal,combrexelle2015bayesian}, the distribution of the log-leaders $\boldsymbol{\ell}_{j}$ can be well-approximated by a multivariate Gaussian distribution with density given by:
\begin{equation}
	\label{eq_log_gaussian2}
	p(\boldsymbol{\ell}_j | \theta_1, \theta_2) \propto 
	{{\big| \,\, \mathbf{\Sigma}_j(\boldsymbol{\theta}) \big|^{-1/2} }}
	{\exp \Big( -\frac{1}{2} \boldsymbol{\ell}_j^T  \mathbf{\Sigma}_j(\boldsymbol{\theta},\norm{\Delta \textbf{n}})^{-1}   \boldsymbol{\ell}_j \Big)}.
\end{equation} 
{Assuming independence between scales, one obtains:}
\begin{equation}
\label{eq_log_gaussian}
     p(\boldsymbol{\ell} | \theta_1, \theta_2) = \prod_{j=j_1}^{j_2} p(\boldsymbol{\ell}_j | \theta_1, \theta_2) ,
\end{equation} %
where $|\textbf{A}|$ denotes the determinant of a matrix $\textbf{A}$, $\boldsymbol{\theta} \triangleq[{\theta}_1, {\theta}_2]^T$, $-c_2 ={\theta}_1 $, ${\theta}_2$ is related to the model adjustment constant $c_2^0$, and $\mathbf{\Sigma}_j(\boldsymbol{\theta},\norm{\Delta \textbf{n}}) \approx  \textrm{Cov}(\boldsymbol{\ell}_{j,\mathbf{n}}, \boldsymbol{\ell}_{j,\mathbf{n} +  \Delta \mathbf{n}})  $ is a covariance model \cite{combrexelle2015bayesian}, defined by the radial symmetric function
\begin{equation}
\label{eq_covariance}
     \mathbf{\Sigma}_j(\boldsymbol{\theta},r) \triangleq {\theta}_1 g_{1}(j,r) +{\theta}_2 g_{2}(j,r),
\end{equation}
where $r \triangleq \norm{\Delta \textbf{n}} $, $g_{1}$ and $g_{2}$ are functions defined as $ g_{1}(j,r) = \max \{ 0, - \ln \big( (r+1)/ (r_j + 1) \big) \} $ and $g_{2}(j,r) =  \max \{ 0, - \ln (r+1) / \ln 4 \}$ with $r_j = \lfloor N_j / 4 \rfloor$ \cite{leon2022bayesian}.

The likelihood \eqref{eq_log_gaussian} is {difficult to be used} in a numerical algorithm but can be efficiently approximated in the spectral domain by formulating an equivalent Fourier-domain data augmented model introduced in \cite{wendt2018multifractal}. This statistical model is based on a Whittle approximation and data augmentation allowing the suitable design of conjugate priors for the parameters ${\theta}_1 , {\theta}_2$ \cite{wendt2018multifractal,combrexelle2015bayesian}. More precisely, \eqref{eq_log_gaussian} is approximated as
\begin{equation}
\label{eq_logleaders_likelihood}
 p(\boldsymbol{\ell}|{\theta}_1 , {\theta}_2) \approx p \big( \textbf{x}, \boldsymbol{\mu}| {\theta}_1 , {\theta}_2  \big), 
\end{equation}
where $p \big( \textbf{x}, \boldsymbol{\mu}| {\theta}_1 , {\theta}_2  \big)$ is given by
\begin{equation}
	\label{eq_data_augmentation_no_K}
	p \big( \textbf{x}, \boldsymbol{\mu}| {\theta}_1 , {\theta}_2  \big) 
	\propto  
	{\theta}_1^{-S} \exp\bigg( -\frac{1}{{\theta}_1} (\textbf{x} - \boldsymbol{\mu})^{H} G^{-1}_{1,s} (\textbf{x} - \boldsymbol{\mu}) \bigg)  {\theta}_2^{-S} \exp\bigg(- \frac{1}{{\theta}_2}  \boldsymbol{\mu}^{H} G^{-1}_{2,s} \boldsymbol{\mu} \bigg), 
\end{equation}
where $\boldsymbol{\mu}$ is a latent vector and $\textbf{x} =[\textbf{x}_{1}^T ,..., \textbf{x}_{S}^T]^T  $ is a subset of Fourier coefficients of $\boldsymbol{\ell}$ \cite{wendt2018multifractal,leon2022bayesian},
\begin{equation}
	\label{eq_x_Fourier_}
	\textbf{x}_{s} \triangleq \textbf{x}(\boldsymbol{\omega}_{s}) = 
	\frac{1}{{N_j}} \sum_{\textbf{n}\in \llbracket - {N_j}, {N_j} \rrbracket^{2}}
	\boldsymbol{\ell}_{j,\mathbf{n}} 
	\exp{(-i \, \mathbf{n}^T \boldsymbol{\omega}_{s})},
\end{equation}
where $ s \in \big\{(j,\textbf{m}): j = \llbracket j_1, j_2 \rrbracket, \textbf{m} \in \Omega^{\dagger}_j \big\}$, $\Omega^{\dagger}_j $ indexes low frequencies (see \cite{wendt2018multifractal,leon2022bayesian} for further details), and $\boldsymbol{\omega}_{s} = 2\pi \mathbf{m}/\sqrt{N_j} $ is the given Fourier frequency, with $(j,\textbf{m}) \in s$. 
Moreover, {the matrices}
\begin{equation}
	\label{eq_functions_G}
	{G}_{i,s} \triangleq
	 \sum_{ \textbf{n}\in \llbracket - {N_j}, {N_j} \rrbracket^{2}} 
	g_{i}(j,\norm{\textbf{n}})  
	\exp{(-i \, \textbf{n}^{T} \boldsymbol{\omega}_{\mathbf{s}}   )},
\end{equation}
for $i=1,2$ are real and strictly positive matrices defined on $s$.
The corresponding spectral covariance of the asymptotic approximation \eqref{eq_data_augmentation_no_K} is expressed as
\begin{equation}
\label{eq_covariance_STG}
    \mathbf{\Xi}_{s} = {\theta}_{1} {G}_{1,s}+{\theta}_{2}{G}_{2,s},
\end{equation}
which leads to standard conditional distributions and fast estimation when used in a Bayesian model \cite{wendt2018multifractal,leon2022bayesian}.

\section{Towards Multifractal-based Image Segmentation in the Fourier Domain}
\label{sect:proposal}

Model \eqref{eq_data_augmentation_no_K} provides an accurate estimation of $c_2$ (recall that $-c_2 ={\theta}_1 $). However, it is not designed for images with several multifractal textures. Here, we propose a new statistical model in the Fourier domain for modeling heterogeneous multifractal images. The limitation of \eqref{eq_data_augmentation_no_K} for computing $c_2$ is its reliance on a regular spatial grid. This prohibits its use for scenarios involving heterogeneous image regions with irregular shapes having different multifractal properties. We therefore develop a spectral likelihood that can handle \RCH{multifractal regions with irregular shapes} \cite{fuentes2007approximate,Guillaumin2022}, resulting in a more flexible model. \RCH{It is important to note that this spectral likelihood is developed by means of the debiased spatial Whittle likelihood from \cite{Guillaumin2022}, used and adapted to address the complexity of computing the likelihood in the Fourier domain for irregularly shaped regions on regularly sampled images. Although the approach could be adapted to handle truly irregularly sampled data, this is beyond the scope of this work.
} %


\subsection{Likelihood on \RCH{Regions With} Irregular \RCH{Shapes}} 
\label{sect_multifractal_for_K} 

Consider an image $\textbf{F}$ that is composed of {regions} with different multifractal spectra. The log-leaders of $\textbf{F}$ can be assumed to be a mixture of $K$ classes denoted as $\mathcal{C}_k$ with different multifractal parameters $\boldsymbol{\theta}_{k} \triangleq [  {\theta}_{1}^{k}, {\theta}_{2}^{k} ]^{T}$, $k=1,...,K$. For segmentation purposes, let $z_{j,n}$ be the label at position $n$ and scale $j$ that maps the observation ${\ell}_{j,n}$ to the class $\mathcal{C}_k$. {The label vector at scale $j$ is denoted as} $\textbf{z}_{j} = [z_{j,1},...,z_{j,N^2_j}]^T$ and ${\textbf{z}} = [\textbf{z}^T_{j_1},...,\textbf{z}^T_{j_2}]^T $. 
Let ${I}_{j,k} = \{ {n} \in {N}^{2}_j \, |  {z}_{j,n} = k\} $ be the subset of sites in $\textbf{z}_{j}$ with label $k$ at scale $j$, and $\mathcal{T}_{j,\mathbf{n}, k}$ its corresponding indicator function:
\begin{equation}
    \mathcal{T}_{j,\mathbf{n},k} = \left\{\begin{matrix}
    1,& \mathbf{n} \in  I_{j,k} \\ 
    0,& \mathrm{otherwise}. 
    \end{matrix} \right.
\end{equation}
The likelihood \eqref{eq_log_gaussian} can then be rewritten as %
\begin{equation} 
\label{eq_LHHH_New}
  p( \boldsymbol{\ell} | \boldsymbol{\Theta}, \textbf{z} ) = \prod_{j=j_1}^{j_2} \prod_{k=1}^{K} 
  p(\ell_{j,\mathbf{n} \in I_{j,k} }|\boldsymbol{\theta}^{T}_K),
\end{equation}
where $p(\ell_{j,\mathbf{n} \in I_{j,k} }|\boldsymbol{\theta}^{T}_K)$ is the likelihood of a stationary zero mean Gaussian process with covariance \eqref{eq_covariance} observed at locations $\mathbf{n} \in I_{j,k}$, and $\boldsymbol{\Theta} =  [\boldsymbol{\theta}^{T}_1,..., \boldsymbol{\theta}^{T}_K]^{T} $ contains the whole set of multifractal parameters.
Just like \eqref{eq_log_gaussian},  \eqref{eq_LHHH_New} is problematic for direct use in practice. In addition, obtaining an approximation similar to \eqref{eq_data_augmentation_no_K} for \eqref{eq_LHHH_New} is less straightforward. 

\subsubsection{Debiased Fourier Coefficients}

Having different multifractality regions implies rewriting \eqref{eq_data_augmentation_no_K} using the coefficients that share the same multifractality, yielding \RCH{an} irregular \RCH{shape} which requires to adjust for bias in the Fourier transform \cite{fuentes2007approximate,Guillaumin2022}.
The log-leader Fourier coefficients are redefined for the $k$-th class label in terms of the irregular \RCH{domain} $\mathcal{T}_{j,{n},k}$ as 
\begin{equation}
    {\textbf{x}}_{s,k} \triangleq 
    \xi_{j,k}
    \sum_{\textbf{n}} 
    \mathcal{T}_{j,\mathbf{n},k}
    \big[ 
    \boldsymbol{\ell}_{j,\textbf{n}} - t_{j,k}
    \big]
    \exp{(-i \, \textbf{n}^{T} \boldsymbol{\omega}_{s}   )},
    \label{eq_fourier_coefficients_s}
\end{equation}
where $\xi_{j,k} = 1/\big({\sum_{\mathbf{n}} \mathcal{T}^{1/2}_{j,\mathbf{n},k}}\big)$ and $t_{j,k} = \xi^{-1}_{j,k}{\sum_{\mathbf{n}}\mathcal{T}_{j,\mathbf{n},k}  \boldsymbol{\ell}_{j,\mathbf{n}} }$ are scaling constants considering the bias introduced by the irregular \RCH{domain} \cite{fuentes2007approximate,Guillaumin2022}.

\subsubsection{Covariance Matrix} The covariance matrix in \eqref{eq_covariance_STG} for the $k$-th class is rewritten as
\begin{align}
   \mathbf{\Xi}_{s,k} \triangleq
    & \mathbf{W}_{s, k} \, \mathbf{\Xi}_{s} ,
    \\ 
    = 
    & 
    {\theta}^k_{1}  \,\mathbf{W}_{j,\textbf{n}, k}  \, {G}_{1,s} 
    +
    {\theta}^k_{2} \, \mathbf{W}_{j,\textbf{n}, k} \, {G}_{2,s} ,
\end{align}
where $\mathbf{W}_{j,\textbf{n}, k}$ is a weight matrix defined as \cite{fuentes2007approximate,Guillaumin2022}:
\begin{equation}
    \mathbf{W}_{j,(n_1, n_2),k} = \xi_{j,k} 
    \sum_{u \in \mathcal{R}_j}
    \sum_{r \in \mathcal{R}_j} 
    \mathcal{T}_{j,(u,r),k}
    \mathcal{T}_{j,(u + n_1, r + n_2),k},
    \label{eq_debiased}
\end{equation}
where $\mathcal{R}_j =  \llbracket - {N_j}, {N_j}-1 \rrbracket^{2}$ is a regular grid. 
The likelihood of the $k$-th set is given by
\begin{equation}
    p({\textbf{x}}_{k} | {\theta}^{k}_1,{\theta}^{k}_2, \textbf{z}) 
    \propto  \prod^{S}_{{s}=1} |{\mathbf{\Xi}}_{s,k}|^{-1/2} 
    \exp 
    \big(-{\textbf{x}}_{{s},k}^{H}  \mathbf{\Xi}_{s,k}^{-1}  {\textbf{x}}_{{s},k}
    \big),\label{eq:likelihood_at_K}
\end{equation}
where ${\textbf{x}}_{k} = [{\textbf{x}}_{1,k},...,{\textbf{x}}_{S,k}]^T$ is the vector of Fourier coefficients of class $k$.

Accordingly, making use of data augmentation as in \eqref{eq_data_augmentation_no_K}, the likelihood \eqref{eq:likelihood_at_K} can be written in a form that factorizes terms with the parameters ${\theta}^{k}_1$ and ${\theta}^{k}_2$, as
\begin{equation}
     \label{eq_data_augmentation_K}
      p \big( {\textbf{x}}_{k} , {\boldsymbol{\mu}}_{k}| {\theta}^{k}_1 , {\theta}^{k}_2 , \textbf{z} \big)   
      \propto 
      \prod^{S}_{{s}=1} ({\theta}^{k}_2)^{-S} \exp\bigg( -\frac{1}{{\theta}^{k}_2}  \boldsymbol{\mu}_{k}^{H} \mathbf{W}_{s,k} G^{-1}_{2,s} \boldsymbol{\mu}_{k} \bigg) 
      ({\theta}^{k}_1)^{-S}    \exp\bigg( -\frac{1}{{\theta}^{k}_1} ({\textbf{x}}_{k} - {\boldsymbol{\mu}}_k)^{H} \mathbf{W}_{s,k} G^{-1}_{1,s} ({\textbf{x}}_{k} - \boldsymbol{\mu}_{k}) \bigg),
\end{equation}
where $\mathbf{W}_{s,k} \triangleq \mathbf{W}_{j,\textbf{n}, k}$, and ${\boldsymbol{\mu}}_{k} = [{{\mu}}_{1,k}^T ,..., {{\mu}}_{S,k}^T]^T $ is the complex-valued latent vector. 
The likelihood for all classes is {finally defined as:} 
\begin{align}
     \label{equ:aproxxLH}
      p \big( {\textbf{x}} , {\boldsymbol{\mu}} | \boldsymbol{\Theta} , \textbf{z} \big)  =      
      \prod_{k=1}^{K} p \big( {\textbf{x}}_{k} , {\boldsymbol{\mu}}_{k}| \boldsymbol{\theta}_k, \textbf{z} \big),
\end{align}
where $\boldsymbol{\theta}_{k} = [  {\theta}_{1}^{k}, {\theta}_{2}^{k} ]^{T}$. %
Finally, as in \eqref{eq_logleaders_likelihood}, we use \eqref{equ:aproxxLH} as an approximation for \eqref{eq_LHHH_New}:  
\begin{align}
\label{equ:aproxxLHbis}
p(\boldsymbol{\ell}|   \boldsymbol{\Theta},\textbf{z}  ) &\approx
 p \big( {\textbf{x}} , {\boldsymbol{\mu}} | \boldsymbol{\Theta} , \textbf{z} \big).
\end{align}

\section{Bayesian Model for Multifractal Segmentation}

With the likelihood \eqref{equ:aproxxLH}, the multifractal parameters and labels corresponding to image regions with different regularities can be estimated using Bayesian estimators. This approach uses Bayesian inference, where prior distributions are assigned to the unknown model parameters, and these parameters are estimated using the resulting posterior distribution.
In this section, the proposed Bayesian model for image segmentation and multifractal parameter estimation is presented.

\begin{figure}[t]
    \centering
    \includegraphics[width=0.9\linewidth]{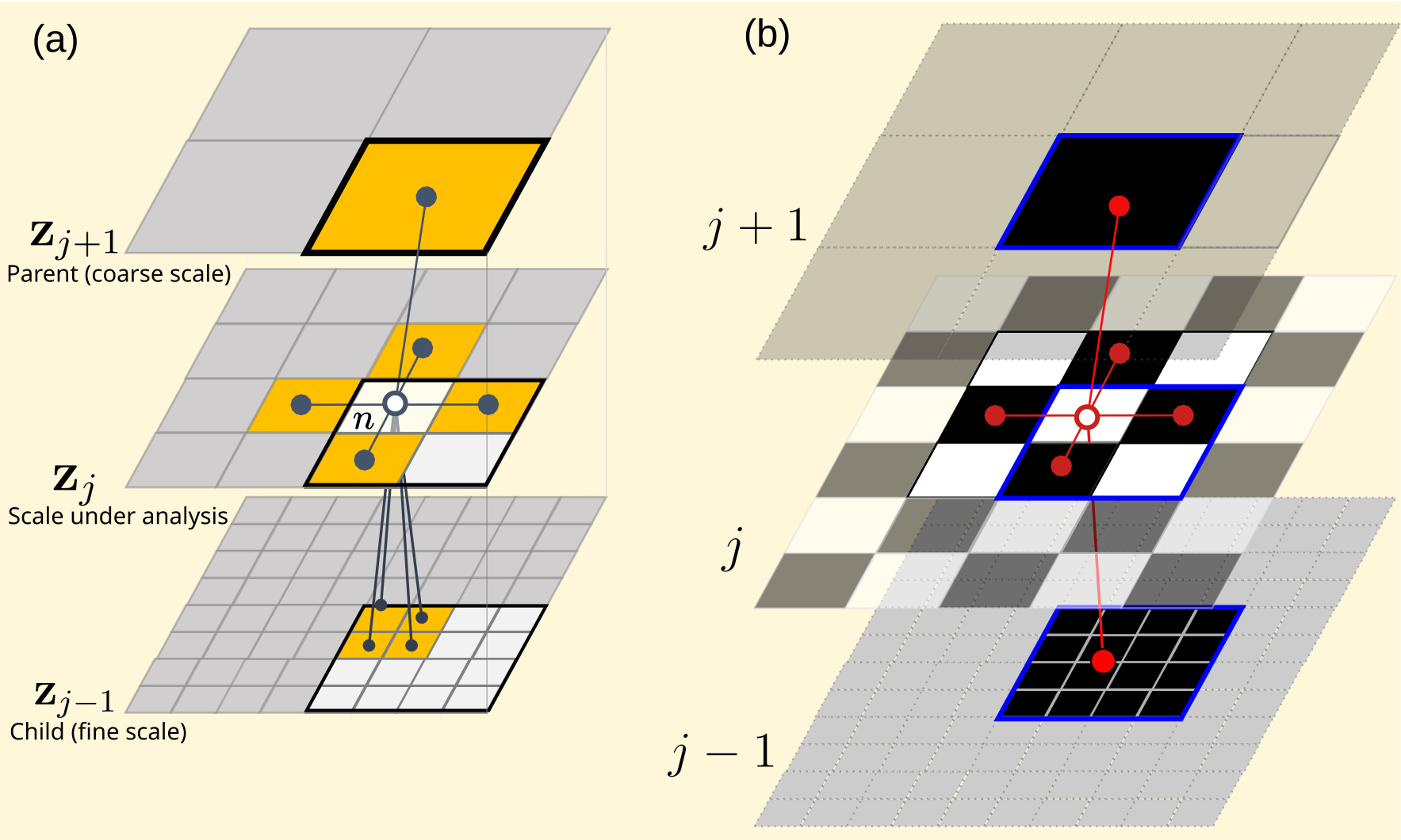}
    \caption{(a) Multiscale label graph structure considered as a prior on the log-leader coefficients. The label at position $n$ (white circle) uses its spatial and scale neighbors (gray dots) to update its value accordingly. (b) Chessboard label decomposition for parallelizing the label sampling in the Potts-Markov random field, where the image at scale $j$ is divided into two independent sets of pixels — white and black, and they are assumed independent. For each scale, pixels are updated using the chessboard scheme considering the spatio-scale connections marked in red. }
    \label{fig:graph}
\end{figure}

\subsubsection{Multiscale Label Prior}

To account for the multiscale nature of the log-leaders, the {label} dependencies within and across scales are modeled via a graph. Specifically, \textit{inter-scale} connections link neighbors across scales, and the \textit{intra-scale} connections model the spatial correlation of labels. Figure \ref{fig:graph}-(a) illustrates the proposed graph structure at a given scale $j$, where the labels to be sampled at position $n$ (white circle) use the values of its neighbors in space and scale (gray dots) to update their values accordingly. %
Formally, the corresponding multiscale prior for $\textbf{z}_{j} \in \{1,...,K\}^{n}$ at scale $j$ is defined as a modified version of the classical Potts-Markov field \cite{wu1982potts},
\begin{equation} 
\label{eq_proposed_prior}
    p\big(\textbf{z}_j  | \textbf{z}_{-j}, \beta_s, \beta^{j}_{xy}\big) =  \frac{1}{\tilde{C}}{\,\exp \Big [  \Phi_{\beta_s} (\textbf{z}_{j}, \textbf{z}_{-j}) + \Phi_{\beta^{j}_{xy}} (\textbf{z}_{j}) \Big ]
    } ,
\end{equation}
where $\textbf{z}_{-j}$ denotes the labels $\textbf{z}_{j-1}$, $\textbf{z}_{j+1}$ on adjacent scales (if $j=1$ or $j=J$, then $\textbf{z}_{-j}=\textbf{z}_{2}$ or $\textbf{z}_{-j}=\textbf{z}_{J-1}$, respectively).

The potential for the \textit{spatial} (intra-scale) dependencies is:
\begin{equation}
	\Phi_{\beta^{j}_{xy}} (\textbf{z}_{j}) \triangleq 
	\sum_{n=1}^{N^2_j} 
	\sum_{m\in \mathcal{V}(n)}
	\beta^{j}_{xy} \delta\Big({z}_{j,n} - \textbf{z}_{j,m}\Big) ,
	\label{eq_phi_xy}
\end{equation}
where $\mathcal{V}(n) $ contains the four spatial neighbors of $n$ \cite{pereyra2012segmentation}, as depicted in Fig. \ref{fig:graph}-(a).
The potential for the \textit{scale} (inter-scale) dependencies is defined as:
\begin{equation}
	\Phi_{\beta_s} (\textbf{z}_{j},\textbf{z}_{-j}) \triangleq 
	\sum_{n=1}^{N^2_j} \sum_{p\in D(n)} \beta_s \, \delta\Big({z}_{j,n} - \textbf{z}_{j+1,p}\Big) +  \sum_{q\in U(n)} \beta_s \, \delta \Big({z}_{j,n} - \textbf{z}_{j-1,q}\Big) ,
	\label{eq_phi_s}
\end{equation}
where $D(n)$ and $U(n)$ are multiscale index mapping functions that map the index $n$ to its corresponding parent (\textit{down}) or child (\textit{up}) coefficient at a given scale, respectively (see Fig. \ref{fig:graph}-(a)).
The inter-scale  and spatial granularity coefficients $\beta_s$ and $\beta^{j}_{xy}$, respectively, adjust the degree of homogeneity across scales and in space, and are gathered in the vector $\boldsymbol{\beta} = [\beta_{s},\beta^{1}_{xy},...,\beta^{J}_{xy} ]$.
Finally, $\tilde{C} = C( \beta_s, \beta^{j}_{xy})$ is a partition function defined as
\begin{equation} 
     C( \beta_s, \beta^{j}_{xy}) \triangleq 
     \sum_{\textbf{z}_{j} } \exp \big[ \Phi_{\beta_s} (\textbf{z}^{j},\textbf{z}_{-j}) + \Phi_{\beta^{j}_{xy}} (\textbf{z}_{j})\big],
     \label{eqC}
\end{equation}
which is generally intractable because the number of summation terms grows exponentially with the size of $\textbf{z}_{j}$ \cite{pereyra2013estimating}.

\subsection{Parameter Priors}

\subsubsection{Granularity Parameters $\beta_s$ and $\beta^j_{xy}$}

Fixing a specific value for the parameters $\beta_s$ and $\beta^j_{xy}$ in \eqref{eq_proposed_prior} is a difficult task due to individual image's spatial structure and organization that can change from one image to another. Indeed, small values of $\beta^j_{xy}$ lead to noisy segmentation whereas large ones lead to coarse or over-smoothed regions. Inspired by \cite{pereyra2014maximum,pereyra2013estimating}, the granularity parameters are jointly estimated with the labels and are assigned a uniform prior distribution on the interval $(0,Q)$, i.e., $f(\boldsymbol{\beta}) = \mathcal{U}_{(0,Q)} (\boldsymbol{\beta})$, where $\boldsymbol{\beta} = [\beta_s , \beta^1_{xy},...,\beta^J_{xy}] $ and 
$Q$ is a given constant related to the number of classes $K$.

\subsubsection{Multifractal Parameter Prior}

The natural conjugate priors for the parameters $\boldsymbol{\theta}^k_i$ are inverse-gamma distributions $\mathcal{IG}(\alpha_{i,k},\gamma_{i,k} ) $ \cite{wendt2018multifractal,combrexelle2015bayesian}, with fixed constants $\alpha_{i,k},\gamma_{i,k}$ $(i=1,2)$ defined, here, for each multifractal region $k = 1,...,K$.


\subsection{Posterior Distribution}

\begin{figure}[t]
    \centering
    \includegraphics[width=0.8\linewidth]{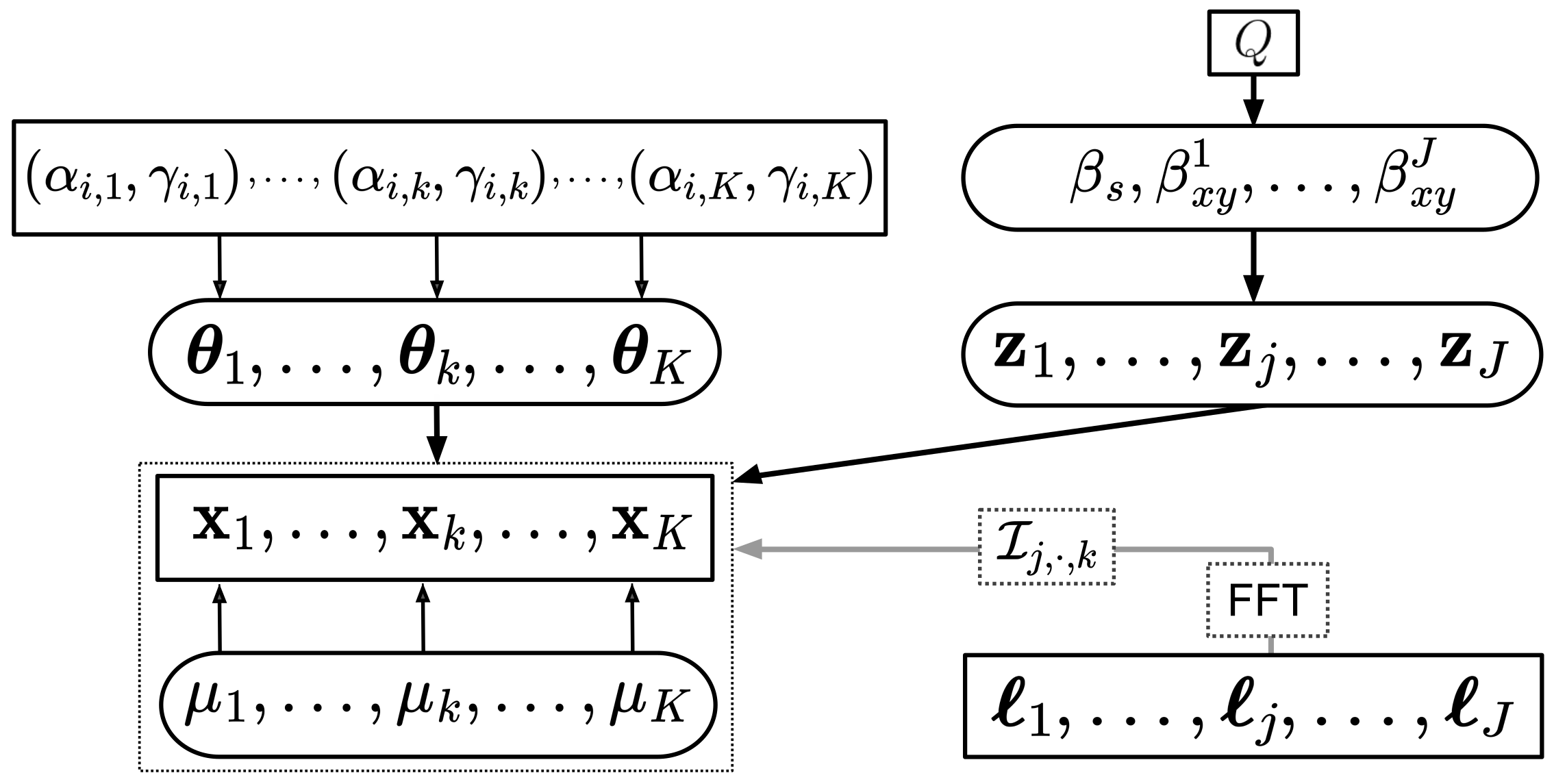}
    \caption{Directed acyclic graph (DAG) of the Bayesian model: the ellipse nodes show the estimated quantities and the square ones, the fixed ones. The FFT transform links the $\boldsymbol{\ell}$ parameters to $\textbf{x}$ and $\boldsymbol{\mu}$.}
    \label{fig:dag_fig}
\end{figure}

The posterior distribution {of} the model parameters {depends on} the spectral likelihood \eqref{equ:aproxxLH} and the priors as {follows:}
\begin{equation}
\label{eq_posterior_augmented_all}
    p(\boldsymbol{\Theta},{\boldsymbol{\mu}}, \textbf{z}, \boldsymbol{\beta}  | {\textbf{x}}) \propto  \,\, p \big( {\textbf{x}} , {\boldsymbol{\mu}} | \boldsymbol{\Theta} , \textbf{z} \big) \,\,
    p(\boldsymbol{\Theta})\,\,
    p(\textbf{z} | \boldsymbol{\beta} ) \,\, p(\boldsymbol{\beta}).
\end{equation}
The Bayesian model is summarized in Fig. \ref{fig:dag_fig}. Note that given the complexity of the posterior distribution \eqref{eq_posterior_augmented_all}, closed-form expressions for the estimators of the unknown parameters $\boldsymbol{\Theta}$, $\textbf{z}$ and $\boldsymbol{\beta}$ cannot be obtained. As an alternative, it is possible to generate samples that are asymptotically distributed according to the target distribution \eqref{eq_posterior_augmented_all} \cite{pereyra2013estimating}. These samples {are then} used to approximate Bayesian estimators of the unknown model parameters.

\section{Sampling Procedure}

This section describes a Gibbs sampler that draws samples distributed according to the posterior distribution \eqref{eq_posterior_augmented_all} by making use of its conditional distributions.

\subsection{{Conditional Probability $p(\mathbf{z} | {\mathbf{x}},{\boldsymbol{\mu}}, \boldsymbol{\Theta},\boldsymbol{\beta})$}}

The conditional distribution for all labels $\textbf{z}$ is
$$
p(\textbf{z} |  {\textbf{x}},{\boldsymbol{\mu}}, \boldsymbol{\Theta},\boldsymbol{\beta})
\propto
p \big({\textbf{x}} , {\boldsymbol{\mu}} | \boldsymbol{\Theta} , \textbf{z} \big) \,\,
    p(\textbf{z} | \boldsymbol{\beta} ).
$$
{The vector} $\textbf{z}$ cannot be sampled jointly since its entries are correlated by the Potts prior.
The conditional distribution of a label $z_{j,n}$, $1<j<J$, given all the other labels $\textbf{z}_{\backslash (j,n)}$, is
\begin{equation}
    \label{eq:multiresolution_graph_0}
    {z}_{j,{n}} \, | \, \textbf{z}_{\backslash (j,n)},
{\textbf{x}},{\boldsymbol{\mu}}, \boldsymbol{\Theta},\boldsymbol{\beta} 
    \propto   \!
 p({z}_{j,{n}}| \textbf{z}_{\backslash (j,n)},\boldsymbol{\beta})  \, p \big( {\textbf{x}} , {\boldsymbol{\mu}} | \boldsymbol{\Theta} , \textbf{z}_{\backslash (j,n)} \big)
\end{equation}
with
$$ 
    \label{eq:multiresolution_graph}
        p({z}_{j,{n}}| \textbf{z}_{\backslash (j,n)},\boldsymbol{\beta}) \triangleq  
         \exp \Bigg[  \sum_{m \in \mathcal{V}({n})} \beta^{j}_{xy} \, \delta \Big({z}_{j,n} - {z}_{j,m} \Big) + 
         \sum_{p \in D(n)} \beta_s \, \delta \Big({z}_{j,n} - {z}_{j+1,p}\Big)
         + \sum_{q \in U(n)} \beta_s \, \delta \Big({z}_{j,n} - {z}_{j-1,q} \Big)\Bigg].
$$
Using \eqref{eq_logleaders_likelihood} and dropping all the terms independent of ${z}_{j,{n}}$, {the following result is obtained:}
$$
p \big( {\textbf{x}} , {\boldsymbol{\mu}} | \boldsymbol{\Theta} , \textbf{z}_{\backslash (j,n)} \big) \propto p(\ell_{j,n} | {z}_{j,{n}}, \boldsymbol{\Theta}  ),
$$
and the marginal density of $\ell_{j,n} $ is:
\begin{equation}\label{eq_LHHH_New_part2}
	p(\ell_{j,n} | {z}_{j,{n}}=k, \boldsymbol{\Theta}  )\sim  \mathcal{N}(0,\sigma^2_j(\boldsymbol{\theta}_k)),
\end{equation}
where $\mathcal{N}(0,\sigma^2)$ denotes a zero-mean Gaussian distribution with variance $\sigma^2$, and $\sigma^2_j(\boldsymbol{\theta}_k) = {{\theta}_2^k + {\theta}_1^k\log 2^j }$.

The multiscale labels are efficiently sampled using a checker-board parallelized strategy \cite{gonzalez2011parallel,pereyra2013estimating}, here adapted to work at multiple resolutions for the multiscale label prior, as illustrated in Fig.\ref{fig:graph}-(b). The proposed strategy decomposes the set of labels at a given scale $j$ into two independent subsets, containing odd and even positions (represented in black and white in Fig.\ref{fig:graph}-(b)). All labels within a subset are conditionally independent and can thus be sampled in parallel, with the labels of the other subset fixed, allowing odd and even labels to be sampled alternatively.

\subsection{Sampling the Granularity Parameters $\beta_s, \beta^j_{xy} $ } \label{section_conditional_beta}

It has been shown in \cite{pereyra2013estimating} that it is possible to approximate the conditional density $f(\boldsymbol{\beta} | \mathbf{\Theta}, \mathbf{x}, \boldsymbol{\mu}, \mathbf{z} ) = f(\boldsymbol{\beta} | \mathbf{z} ) $ by considering an auxiliary vector $\mathbf{w}$ that verifies $\mathbf{\Phi}(\mathbf{z}) = \mathbf{\Phi} (\mathbf{w}) $, where $\mathbf{\Phi}(\mathbf{z})$ is a tractable sufficient statistic. Then, by considering the Euclidean distance $\mathbf{\Phi}(\mathbf{z}) - \mathbf{\Phi} (\mathbf{w}) $, an approximation of the gradient can be computed. In practice, $\mathbf{w}$ is initialized with $\mathbf{z}$, and then updated with few Gibbs moves associated with the target density $f({\mathbf{z}}| \boldsymbol{\beta}  )$ (for more details see \cite{pereyra2014maximum,pereyra2013estimating}). This approximation of the gradient results in values around the neighborhood of $\boldsymbol{\beta}$ when choosing the correct statistic $\mathbf{\Phi}(\mathbf{z})$.
Inspired by \cite{pereyra2013estimating}, we propose the statistic $\mathbf{\Phi}(\mathbf{z})= \sum_{j=j_1}^{j_2} \Phi_{\beta^{j}_{xy}} (\textbf{z}_{j}) + \Phi_{\beta_s} (\textbf{z}_{j}, \textbf{z}_{-j}) $, for the scale granularity, where $\Phi_{\beta^{j}_{xy}}$ and $\Phi_{\beta_s}$ are defined in \eqref{eq_phi_xy} and \eqref{eq_phi_s}, and, for the spatial parameters, we use the statistic $\Phi_{\beta^{j}_{xy}}(\mathbf{z}_j)$, $j=j_1,...,j_2$.
A truncation operation $T_{[0,Q]}({x})$ is introduced to restrict the parameters to belong to the interval $(0,Q)$, $Q\in\mathbb{N}$.
Following this heuristic but effective strategy, the granularity parameters are sampled as summarized in Algorithm \ref{alg:beta}. 
Intuitively, the sampler performs gradient descent to update the $\boldsymbol{\beta}$ parameters. It first generates auxiliary variables $\textbf{w}_{j}$ using Gibbs sampling. Then, the differences between the statistics evaluated in $\textbf{z}$ and $\textbf{w}$ are computed to approximate the intractable gradient, which are {used} to update the values of $\boldsymbol{\beta}$.

\subsection{Conditional Probability $f(\mathbf{\Theta}|\mathbf{z},\boldsymbol{\beta}, {\mathbf{x}} , {\boldsymbol{\mu}})$} 

The conditional distributions of the multifractal parameters contained in $\mathbf{\Theta}$ and the latent variables in ${\boldsymbol{\mu}}$ are:
\begin{align}
    {\boldsymbol{\mu}}_k| & {\textbf{x}}_k, {\theta}^{k}_1,{\theta}^{k}_2 \sim \mathcal{CN}
    \Big(
    {\theta}^{k}_1  \tilde{G}_{1,k}  {\textbf{x}}_k , 
    \Big(
    ({\theta}^{k}_1 \tilde{G}_{1,k})^{-1} 
    + ({\theta}^{k}_2 \tilde{G}_{2,k})^{-1} \Big) 
    \Big), \notag
    \\
    {\theta}^{k}_1 | & {\textbf{x}}_k, {\boldsymbol{\mu}}_k,  {\theta}^{k}_2 \sim 
    \mathcal{IG}
    \Big(
    \alpha_{1,k} + S , \gamma_{1,k} + \norm{{\textbf{x}}_k - {\boldsymbol{\mu}}_k}_{\tilde{G}_{1,k}} 
    \Big),\notag
    \\
    {\theta}^{k}_2 | & {\textbf{x}}_k,{\boldsymbol{\mu}}_k,{\theta}^{k}_1  \sim 
    \mathcal{IG}
    \Big(\alpha_{2,k} + S, \gamma_{2,k} +\norm{ {\boldsymbol{\mu}}_k }_{\tilde{G}_{2,k}}
    \Big) ,
    \label{eq_conditional_MFA_parameters}
\end{align}
where $\norm{\mathbf{y}}_{\mathbf{G}} \triangleq \mathbf{y}^H \mathbf{G} \mathbf{y}$, $\tilde{G}_{i,k} = \mathbf{W}_{\mathbf{s}, k} G_{i,\mathbf{s}}^{-1}$.

{\footnotesize \begin{algorithm}[t]
	\DontPrintSemicolon
	
	\SetKwFunction{algo}{}\SetKwFunction{proc}{proc}
	\SetKwProg{myalg}{Function}{\color{white}}{\color{white}}

    \KwInput{$\mathbf{z}^{(t)},\boldsymbol{\beta}^{(t)}:$ current label and granularity vector
    \\ \hspace{0.9cm} $Q:$ constant (default $Q=10$)
    \\ \hspace{0.9cm} $t, V:$ current iteration, number of iterations
    }

 	{   
    Set $\mathbf{z}\leftarrow \mathbf{z}^{(t)}$, $[ \beta_s , \beta^{j_1}_{xy},...,\beta^{j_2}_{xy} ] \leftarrow\boldsymbol{\beta}^{(t)}$
    
    \For{r = 1 to V}
    {   
    
    ${\textbf{w}}_{j}\leftarrow$ draw Gibbs moves of ${\textbf{z}}_{j}$ via \eqref{eq:multiresolution_graph_0}, for $j=j_1,...,j_2$

    Set $\textbf{w} = [\textbf{w}^{T}_{j_1},...,\textbf{w}^{T}_{j_2}]^{T}$

    $\eta = 10 \, (t+r-1)^{-3/4} \big(\sum_{j=j_1}^{j_2} N_{j}\big)^{-1} $
    
    Compute $$(\beta_s)_{(r+1)} = T_{[0,Q]} \Big( (\beta_s)_{(r)} + \eta \big(\mathbf{\Phi}({\textbf{z}}) - {\mathbf{\Phi}}({\textbf{w}}) \big)\Big)$$

    \For{$j=j_1$ to $j_2$}{

    $\mathbf{b}_j = \Phi_{\beta^{j}_{xy}} (\textbf{z}_{j})  - \Phi_{\beta^{j}_{xy}}({\textbf{w}}_{j})$
    
    Compute  $(\beta^{j}_{xy})_{(r+1)} = T_{[0,Q]} \Big( (\beta^{j}_{xy})_{(r)} + \eta \mathbf{b}_j  \Big)$   
    }
    }    
	\KwRet $\boldsymbol{\beta} =  [ \beta_s , \beta^{j_1}_{xy},...,\beta^{j_2}_{xy} ] $}
	\setcounter{AlgoLine}{0}
	\caption{Multiscale granularity parameter sampling}\label{alg:beta}
\end{algorithm}}

\subsection{Gibbs Sampler and Bayesian Estimators}
\label{Bayesian_estimators}

Algorithm \ref{alg:two} presents the procedure for jointly sampling the labels, granularity coefficient, and multifractal parameters. The inputs of the algorithm are the image and hyperparameters assumed to be known beforehand: number of regions $K$, constants $\alpha_{1,k}$, $\alpha_{2,k}$,  total number of generated samples $N_m$, number of samples of the burn-in period $N_b$. The output of the Gibbs sampler are used to build estimators of the parameters and labels. More precisely, the marginal posterior mean (aka minimum mean square error (MMSE)) estimator is used for the multifractal parameters $\boldsymbol{\theta}^{k}_i$, $i=1,2$:
\begin{equation}
    \big(\hat{\boldsymbol{\theta}}^{k}_i \big)^{{\text{\tiny MMSE}}} \approx \frac{1}{( N_m - N_b )} \, \sum_{t=N_{b}+1}^{N_{m}} \big[\boldsymbol{\theta}^{k}_i \big]^{(t)}. 
    \label{mfa_parameter_estimation_}
\end{equation} 
On the other hand, due to the discrete nature of the labels, we propose to estimate them using the maximum a-posteriori (MAP) estimator:

\begin{equation}
    \hat{{z}}_{j,n}^{{\text{\tiny MAP}}} = \underset{z_{j,n} \in \{1,...,K \}}{\mathrm{arg\,max}}  \Big\{ p(z_{j,n} | \textbf{z}_{\backslash (j,n)},{\textbf{x}},{\boldsymbol{\mu}}, \boldsymbol{\Theta},\boldsymbol{\beta}) \Big\},
\end{equation}
for $j = j_1,...,j_2$.
Note that the computed labels are downsampled by a factor $2^{j_1}$ compared to the input image. In practice, the labels computed at the finest scale $\hat{\mathbf{z}}_{j_1} \in \mathbb{R}^{N_1 \times N_1}$ are interpolated (using, e.g, nearest neighbors) up to $N\times N $ to obtain labels for each pixel of the image, see \cite{choi2001multiscale} for alternative strategies. 
Finally, the updates for the granularity parameter vector $\boldsymbol{\beta}$ are performed only during the burn-in period of the sampler, as suggested in \cite{pereyra2014maximum}, in order to reduce the computational complexity. After the burn-in period, $\boldsymbol{\beta}$ is fixed to its current value.

{\small\begin{algorithm}[t]
	
	\DontPrintSemicolon
	\caption{Gibbs sampler}\label{alg:two}

    \KwInput{$\textbf{F}, K:$ input image, number of regions		
    \\ \hspace{0.9cm} $j_1,j_2:$ initial and final scale
    \\ \hspace{0.9cm} $N_m, N_b:$ number of iterations, burn-in period
    \\ \hspace{0.9cm} $V:$ number of iterations for $\boldsymbol{\beta}$ (default $V=2$)
    }
    
    {Initialization:} $ \textbf{z}^{(0)}$, $\boldsymbol{\Theta}^{(0)}$, $\boldsymbol{\beta}^{(0)}, \boldsymbol{\mu}^{(0)} $. 
    
    Compute $J$ log-leaders $\boldsymbol{\ell}$ from $\textbf{F}$ (section \ref{subsection:Fourier-based_likelihood}).\;
        
    \For{$t = 1$ to $N_m$}
    { 
    {Compute Fourier coefficients $\textbf{x} = [\textbf{x}^{T}_1,...,\textbf{x}^{T}_K]^{T}$ using $\textbf{z}^{(t-1)}$ via \eqref{eq_fourier_coefficients_s}}.
    
    Sample $\textbf{z}^{(t)}\sim p\Big[ \textbf{z}  | \boldsymbol{\Theta}^{(t-1)},  \textbf{z}^{(t-1)}, \boldsymbol{\beta}^{(t-1)},\textbf{x},\boldsymbol{\mu}
    \Big]$ via \eqref{eq:multiresolution_graph_0}. 
    
    Sample $\boldsymbol{\Theta}^{(t)} \! \sim \! f\big(\boldsymbol{\Theta} | \boldsymbol{\Theta}^{(t-1)}, \! \textbf{z}^{(t)},\boldsymbol{\beta}^{(t-1)}  \!
    , {\textbf{x}},\boldsymbol{\mu}
    \big)$, and $\boldsymbol{\mu}$ via \eqref{eq_conditional_MFA_parameters}. 

    \eIf{$t < N_b$}
    {
        Generate $\boldsymbol{\beta}^{(t)} \sim f\big(\boldsymbol{\beta}| \boldsymbol{\beta}^{(t-1)},  \textbf{z}^{(t)}\big)$ following Alg. \ref{alg:beta}.  
    }{
    $\boldsymbol{\beta}^{(t)} \leftarrow \boldsymbol{\beta}^{(t-1)}$.
    }
    }
    
\end{algorithm}}

\subsection{Label Initialization} 
\label{sec_initialization}

An effective approach for initializing labels consists of estimating the parameters $(\theta_1, \theta_2)$ locally from the log-leaders corresponding to overlapped patches using the regression \eqref{eq_linear_fit_c2}, where local multifractal estimates are computed on $16\times 16$ overlapping patches (with $75\%$ overlap).
The patch-wise estimates for $\theta_1$  are used to obtain an initial guess of the labels by $k$-means clustering.

\subsection{Computational complexity} 
\label{sec_complexity}

The computational complexity of the proposed approach is dominated by that of the Fourier transforms. Since $N_j\propto N$ for all $j$ {(where $\propto$ means ``proportional to'')}, the {estimation} of the multifractal parameter requires $\mathcal{O}( N^2 \log N )$ operations, performed $K$ times. 
The cost of constructing the debiasing matrix in \eqref{eq_debiased} is $\mathcal{O}(N^2)$.
The multiscale Potts model update also requires $ \mathcal{O}\big(N^2 \big)$ operations to compute the probability of wavelet leaders to belong to each class. 
Consequently, the final complexity of the proposed method is $\mathcal{O}\big( K N^2 \log N \big) $. \RCHH{Although demanding for high-resolution images, computational efficiency can be improved by implementing the debiasing matrix at a lower level (e.g., C/C++) and parallelizing the multiscale Potts updates on a GPU.}

\section{Numerical Experiments}

The proposed joint Bayesian segmentation and parameter estimation approach is first evaluated on synthetic 2D images constructed using multifractal random walks (MRWs) by means of Monte Carlo simulations\footnote{\RCH{The source code is available at \url{https://github.com/kareth-leon/MFA_Seg}}}. Some results obtained using real-world data are then presented. MRWs are non-Gaussian processes with stationary increments and are constructed in order to have multifractal properties mimicking those of Mandelbrot’s log-normal cascades, 
see \cite{robert2010gaussian} for details.

\subsection{Monte Carlo Simulations}

\subsubsection{Synthetic Data}

To validate the proposed method under controlled conditions, two scenarios with known ground-truth labels and multifractal parameters were considered. 
Specifically, the first scenario has $K=2$ different multifractal regions: a background with multifractality parameter $\theta^1_1 = -0.02$ and a centered disk with parameter $\theta^2_1 =-0.08$, whereas the second scenario has $K=3$ regions: a background with $\theta^1_1 = -0.02$ and two disks with parameters $\theta^2_1 =-0.08$ and $\theta^3_1 =-0.16$. 
The experiments were run on 100 independent realizations of {MRWs} with these parameters, where the computation time for one realization is in average 58 seconds\footnote{\RCH{The simulations were conducted in a desktop architecture with an Intel Corei7, 16 GB RAM, using MATLAB R2024b.}}. 
An illustration of a single realization of an image for each scenario and their corresponding true masks is displayed in Figs. \ref{fig:result_MRW_all_methods} (a) and (b), respectively.

\subsubsection{Performance Measure}

The performance of the multifractality parameter estimation is quantified using the mean, the standard deviation (STD) and the root-mean square error (\textsc{RMSE}) of the estimates. 
The label estimation performance is assessed using the dice score coefficient (DSC) and the segmentation error (aka misclassification error), which are defined in terms of true positive (TP), true negative (TN), false positive (FP), and false negative (FN): 
DSC (aka F1-score) is defined on each region as $\textsc{DSC = 2 TP/ (2 TP + FP + FN)}$. The higher the DSC score, the better the segmentation result. The segmentation error is defined by $\textsc{Error = (FP+FN)/(TP+TN+FP+FN)}$. The lower the error, the better the performance.

\subsubsection{Parameter Settings}

The Daubechies mother wavelet with $ N_{\psi} = 1$ vanishing moments is used in the 2D DWT. The hyperparameters of the independent priors for the multifractal parameters were set to  $(\alpha_{i,k},\gamma_{i,k})=(10^{-3},10^{-3}) $ \cite{wendt2018multifractal}. \RCH{In practice, these constants are fixed to small values, ensuring that they are close to non-informative Jeffreys priors \cite{wendt2018multifractal,pereyra2013estimating}}. The set of scales used for the experiment with $K=2$ classes was set to $(j_1,j_2)=(1,3)$, whereas $(j_1,j_2)=(1,2)$ for $K=3$. \RCH{Note that $j_2$ is limited by image dimensions ($j_2 < \log_2 N$), nonetheless, in this study, $j_2$ was set between 2 and 3, as the coarsest scales were found to retain meaningful pixel-level information.}
The number of iterations for all the experiments was set to $N_m = 300$, with a burn-in period of $N_b = 30${, since it showed to be sufficient for sampling convergence (see next subsection for convergence details).}

\begin{figure*}[t]
    \centering
    \includegraphics[width=1\linewidth]{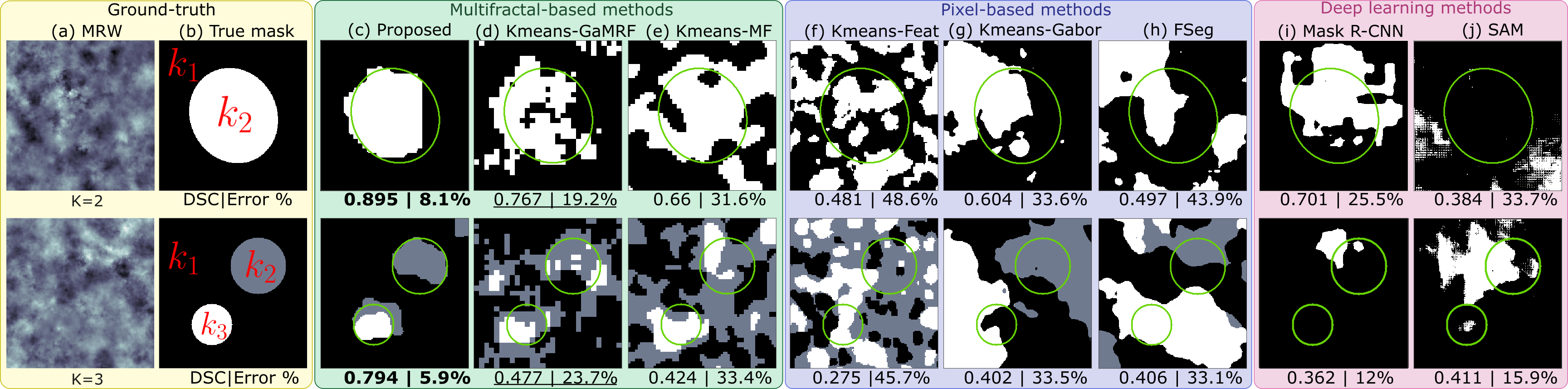}
    \caption{Single MRW realization for $K=2$ (top row) and $K=3$ (bottom row) and their segmentation labels obtained from the different compared methods: (a) MRW image generated using the true labels in (b). The proposed segmentation (c) is compared against methods using multifractal parameters (d) and (e), framed in the green square; classical pixel-based segmentation methods: (f) \textsc{Kmeans-Feat}, (g) \textsc{Kmeans-Gabor}, (h) \textsc{FSeg}, framed in the blue square; and deep-based segmentation approaches: (i) Mask R-CNN, and (j) \textsc{SAM}, framed in the rose square. The average DSC and error percentage of each results are shown in the bottom, where the scores obtained by the proposed method (in bold) are superior regarding the others.
    }
    \label{fig:result_MRW_all_methods}
\end{figure*}


\subsection{Estimation Performance}

\subsubsection{Multifractal Parameter Estimation on Irregular \RCH{Shapes}}

{The proposed approach is first evaluated in a setting where the labels are known and fixed {with $K=2$ classes}, in order to assess the estimation performance of the multifractal parameter on an irregular {domain}. Furthermore, to illustrate the impact of the multifractality parameter value on the image segmentation, the multifractality value in the disk of the MRWs was varied in a range of weak (close to zero) to strong values $\theta^2_1 \in \{-0.005, -0.08, -0.2 \}$, fixing the background parameter to $\theta^1_1 = -0.02$.  
Results are reported in Table \ref{table_parameter_all2}. They indicate that the proposed approach yields satisfactory performance (small RMSE values), comparable to that obtained for regular \RCH{shapes} \cite{wendt2018multifractal,combrexelle2015bayesian}.}

\subsubsection{Multifractal Parameter and Label Estimation}

Table \ref{table_parameter_all3} shows the performance of the multifractal parameter estimation when the labels and granularity parameters are also unknown and estimated jointly, following the previously described setting. 
The results show that the RMSE values slightly increase, due to the larger number of unknowns to be estimated, but that the overall performance remains comparable to Table \ref{table_parameter_all2}.
Table \ref{table_label_all4} reports the DSC values and the segmentation errors for the two experiments, allowing the label estimation performance to be evaluated.  
The results indicate that the segmentation error decreases as the difference in the multifractal parameter between the regions increases. This behavior is indeed expected, since the increased parameter contrast should facilitate segmentation. Overall, the segmentation performance is very satisfactory, except for the extreme case $\theta^2_{1}= -0.005$ that corresponds to nearly no multifractality.

\subsubsection{Sampler Convergence}
\begin{figure}[t]
    \centering
    \includegraphics[width=1\linewidth]{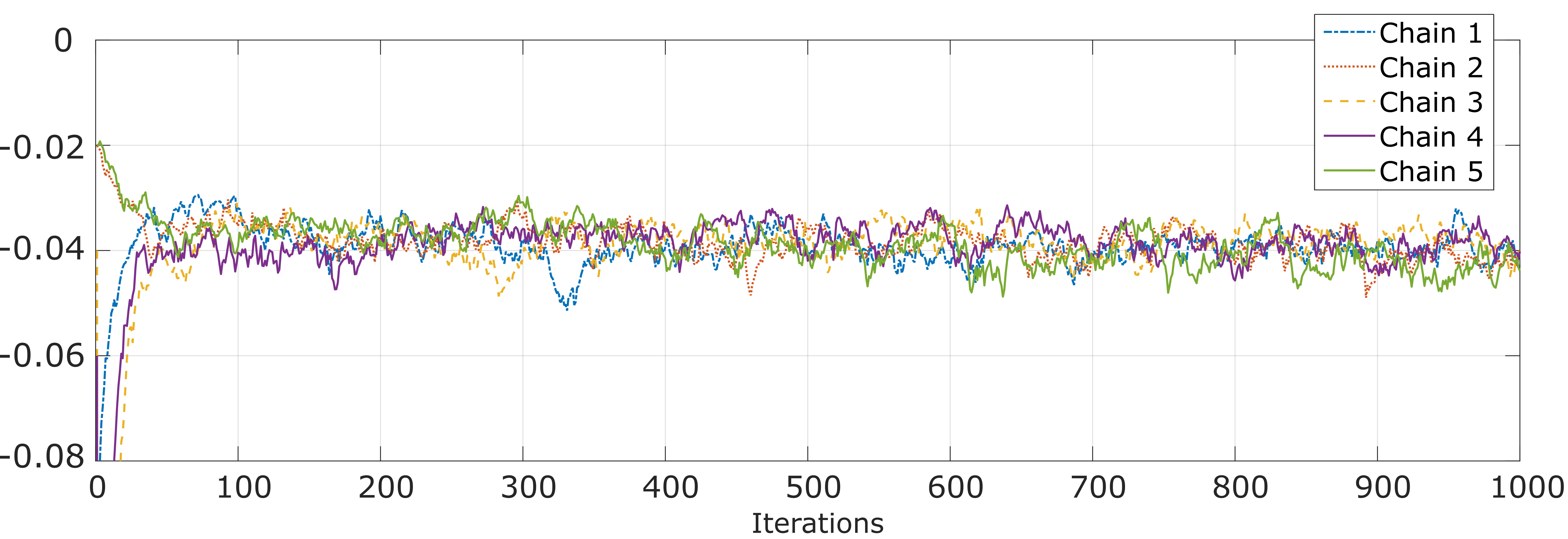}
    \caption{Convergence assessment with five Markov chains with different initializations.}
    \label{fig:conv} 
\end{figure}

\RCH{To evaluate the convergence of the proposed method, we use the potential scale reduction factor (PSRF), a convergence diagnostic for Monte Carlo simulations \cite{gelman1992inference}. It compares the variance within the chain to the variance between chains in multiple chains, where PSRF $< 1.2 $ indicates good convergence of the sampler (see \cite{gelman1992inference,dobigeon2007joint} for further details).}
\RCH{Any parameter sampled in the Monte Carlo simulations can serve as a reference to monitor convergence. For illustration purposes, we chose to monitor the multifractal value of the background of Fig. \ref{fig:result_MRW_all_methods}-a, $\theta^1_1 = -0.02$, during 1000 iterations. Figure \ref{fig:conv} shows five chains of the parameter with different initial values. It can be seen that the chains converge to similar values, showing that the method converges even when changing the initial value of the chain. The PSRF value obtained for this parameter with $1000$ Monte Carlo runs is $1.0009$, exhibiting the convergence of the sampler. }

\renewcommand{\arraystretch}{1}
\setlength{\tabcolsep}{18pt}
\begin{table}[t]
\centering
\caption{Performance of the multifractal parameter estimation $\big(\tilde{\boldsymbol{\theta}}^{k}_1 \big)^{{\text{\tiny MMSE}}}$ when the labels are fixed and known.}
\small
\begin{tabular}{c|c|c|c}
\hline
$k$ & True $\boldsymbol{\theta}^{k}_1$  & \textsc{MEAN (STD)} &RMSE   \\\hline \hline 
1 & $ -0.02$                         & $-0.032\,  (0.005) $     & $0.013$                       \\
2 & $-0.005$                         & $-0.014\, (0.003)$ & $0.010$                        \\\hline \hline 
1 & $-0.02$                          & $-0.034\,(0.005)$ & $0.015$                        \\
2 & $-0.08$                          & $-0.087\, (0.016)$      & $0.018$                        \\\hline \hline 
1 & $-0.02$                          & $-0.034\,   (0.004) $    & $0.015$                       \\
2 & $-0.2$                           & $-0.184\,   (0.028) $    & $0.032$                        \\\hline 
\end{tabular}
\label{table_parameter_all2}
\end{table}

\renewcommand{\arraystretch}{1}
\setlength{\tabcolsep}{16pt}
\begin{table}[t]
\centering
\caption{Performance of the multifractal parameter estimation $\big(\tilde{\boldsymbol{\theta}}^{k}_1 \big)^{{\text{\tiny MMSE}}}$ when labels and parameters are jointly estimated. } %
\small
\begin{tabular}{c|c|c|c}
\hline 
$k$ & True $\boldsymbol{\theta}^{k}_1$ & \textsc{MEAN (STD)} & RMSE      \\\hline \hline 
1 & $ -0.02$                         & $-0.041   \,    (0.014) $      & $0.025$     \\
2 & $-0.005$                         & $-0.023   \,     (0.013 ) $    & $0.022$     \\\hline \hline 
1 & $-0.02$                          & $-0.033   \,    (0.006)$      & $0.014$     \\
2 & $-0.08$                          & $-0.113    \,    (0.025) $     & $0.041$     \\\hline \hline 
1 & $-0.02$                          & $-0.037    \,   (0.011)$      & $0.020$     \\
2 & $-0.2$                           & $-0.238    \,   ( 0.058)$      & $0.069$     \\\hline 
\end{tabular}
\label{table_parameter_all3}
\end{table}

\renewcommand{\arraystretch}{1.5}
\setlength{\tabcolsep}{5pt}
\begin{table}[t]
\centering
\caption{Segmentation performance for the labels in terms of DSC and Segmentation Error in the joint parameters and label estimation scenario for 100 realizations by considering different multifractal parameter $\boldsymbol{\theta}^{(k=2)}_1$ in the disk. }
\small
\begin{tabular}{l|c|c|c}
 \hline
 $\big[\boldsymbol{\theta}^{(k=1)}_1 ,\boldsymbol{\theta}^{(k=2)}_1 \big]$    & \textsc{DSC}$\uparrow$ $[k=1]$ &   \textsc{DSC}$\uparrow$ $[k=2]$ &   \textsc{Error}$\downarrow$ \%  \\\hline\hline 
$[-0.02, -0.005]$  & $0.649 \,    (0.103) $                   & $0.487   \,  (0.223) $                         & $39.4  \, (8.3) $                         \\\hline 
$[-0.02, -0.08]$ & $ 0.909 \,   (0.034)$                         &$ 0.752   \, (0.117)  $                        & $13.2  \, (4.8)   $                      \\\hline 
$[-0.02, -0.2]$ & $0.940   \,(0.030) $                          &$   0.816    \,(0.140) $                          & $9.0  \, (5.0)   $   \\\hline                    
\end{tabular}
\label{table_label_all4}
\end{table}

\subsection{Comparison to the State of the Art}

The proposed segmentation approach is compared to unsupervised state-of-the-art methods for texture$/$image segmentation, including deep learning methods, classical texture segmentation, and the classical $k$-means clustering method, a benchmark in segmentation, used to estimate labels for different features, including multifractal, pixel intensity, and multiscale features. These methods are summarized below: 

\begin{itemize}[leftmargin=5mm]
    \item The $k$-means on multifractal parameters estimated by regression (\textsc{Kmeans-MF}). Note that this method was used to initialize the suggested method described in \ref{sec_initialization}. 
    
    \item The $k$-means on a localized Bayesian multifractal estimates (\textsc{Kmeans-GaMRF}) \cite{wendt2018multifractal}, {which} makes use of a smoothing gamma Markov random field (GaMRF) prior\footnote{Toolbox available at \url{https://www.irit.fr/~Herwig.Wendt/CIMPA2017.html}}. {The resulting} multifractal estimates {are used as features for $k$-means} segmentation.
    
    \item The $k$-means on standard patch-wise image features (\textsc{Kmeans-Feat}): uses patch-wise features including the mean, standard deviation, minimum and maximum values of patches of size $10\times10$ with $80\%$ overlap.
    
    \item The $k$-means on Gabor filter features (\textsc{Kmeans-Gabor}) \cite{jain1991unsupervised}: a benchmark for classical multiscale features extraction without multifractal model.
    
    \item The factorization-based texture segmentation (\textsc{FSeg}) method \cite{yuan2015factorization}: computes multiscale Gabor coefficients followed by a feature selection procedure by a matrix factorization step. This method is selected given its multiscale nature for segmentation. The hyperparameters {were adjusted} as in  \cite{yuan2015factorization}.

    \item The segment anything (\textsc{SAM}) method \cite{kirillov2023segment}: it relies on a \textit{promptable} segmentation that takes as inputs the image and a handcrafted reference (such as points, boxes or mask logits), and returns a segmentation mask \cite{kirillov2023segment}. The pre-trained SAM model was supervised by providing {the} spatial coordinates of two pixels randomly selected from each region.

    \item The \textsc{Mask R-CNN} architecture \cite{he2017mask}: a pre-trained convolutional neural network for instance segmentation that generates a pixel-level binary mask for each detected object in the given image. A threshold of 0.98 was applied to the network's output masks for the detected objects. 
\end{itemize}

\noindent The \textsc{SAM} and \textsc{Mask R-CNN} models were selected since they are outstanding methods of the state of the art in deep-based image/instance segmentation. \RCH{Additionally, note that the \textsc{SAM} model relies on multiple transformers and attention modules that model long-range dependencies to handle global context and complex scenes \cite{kirillov2023segment}. }

Figure \ref{fig:result_MRW_all_methods} displays the segmentation results obtained using the different methods for a single MRW realization, when $K=2$ (top row) and $K=3$ (bottom row). 
The proposed method {yields} an average DSC {close to} $0.85$, significantly outperforming the other methods, and provides visually satisfactory results. All other baselines {lead to} significantly worse results, illustrating the difficulty of segmenting multifractal images. Tables \ref{tab:all_comparison_K2} and \ref{tab:all_comparison_K3_DSC} provide a more quantitative and systematic comparison in terms of average performance for 100 independent realizations of MRW for $K=2$ and $3$. It can be seen that the proposed method significantly outperforms the other methods, showing overall best segmentation scores in terms of \textsc{DSC} and \textsc{Error} by a large margin.

\renewcommand{\arraystretch}{1.2}
\setlength{\tabcolsep}{5pt}
\begin{table}[t]
\caption{Segmentation baseline comparison for the label estimation when \underline{${K=2}$} in terms of average DSC and segmentation Error percentage of {100 realizations}, the standard deviation is shown in parenthesis. The best scores are highlighted in bold. } 
\small
\centering
\begin{tabular}{l|c|c|c}
\hline
  Method                              &  {DSC}$\uparrow$ $[k=1]$  & {DSC}$\uparrow$ $[k=2]$  & \textsc{Error}$\downarrow$ \%  \\\hline\hline
\textsc{Proposed}               & $\textbf{0.909} \, (0.034)$ & $\textbf{0.752} \, (0.117)$ & $\textbf{13.2}  \, ( 4.8)$   \\ \hline
\textsc{Kmeans-GaMRF}            & $0.843 \, ({0.027})$ & $0.567 \, (0.097)     $& $22.8   \,  (3.7)$     \\ \hline
\textsc{Kmeans-MF}              & $0.745 \, (0.079)$ & $0.539 \, (0.111)    $ & $32.2  \,(7.9)$     \\ \hline
\textsc{Kmeans-Feat }           & $0.643 \, (0.071)$ & $0.383 \, (0.088)$ & $44.2   \,(5.7)$ \\ \hline
\textsc{Kmeans-GABOR}           & $0.683 \, (0.084)$ & $0.443 \, (0.126)$ & $39.6    \,(8.4)$ \\ \hline
\textsc{FSeg}                   & $0.637 \, (0.050)$ & $0.416 \,(0.077)$ & $44.6  \, (5.2)$ \\ \hline
\textsc{Mask-RCNN }             & $0.813\, ( 0.042)  $ & $0.300 \,(0.250)    $ & $28.5\, ({0.1})   $ \\ \hline
\textsc{SAM }                   & $0.196 \, (0.215)$ & $0.458 \, ({0.054})$ & $62.7 \,   (8.8)$ \\ \hline
\end{tabular} \label{tab:all_comparison_K2}
\end{table}

\renewcommand{\arraystretch}{1.2}
\setlength{\tabcolsep}{4.5pt}
\begin{table}[h]
\caption{Average DSC for {100 realizations} of the baseline comparison when \underline{${K=3}$}. The standard deviation is shown in parenthesis. 
}
\small
\centering
\begin{tabular}{l|c|c|c}\hline
             & {DSC}$\uparrow$ $[k=1]$  &{DSC}$\uparrow$ $[k=2]$ & {DSC}$\uparrow$ $[k=3]$ \\ \hline\hline
\textsc{Proposed}      & $\textbf{0.846} \, (0.072)$ & $\textbf{0.408} \,  (0.147)$ & $\textbf{0.494} \,  (0.256)$ \\ \hline
\textsc{Kmeans-GaMRF}  & $\textbf{0.846} \, {(0.051)}$ & $0.328 \,  (0.124)$ & $0.439 \,  (0.173)$ \\ \hline
\textsc{Kmeans-MF}     & ${0.669} \, (0.094)$ & $0.230 \,  (0.083)$ & $0.390 \,  (0.147)$ \\ \hline
\textsc{Kmeans-Feat }  & ${0.505} \, (0.150)$ & $0.219 \,  (0.068)$ & $0.139 \,  (0.078)$ \\ \hline
\textsc{Kmeans-GABOR}  & ${0.515} \, (0.091)$ & $0.353 \,  (0.102)$ & $0.210 \,  (0.119)$ \\ \hline
\textsc{FSeg}          & ${0.557} \, (0.118)$ & $0.322 \,  (0.113)$ & $0.198 \,  (0.141)$ \\ \hline
\textsc{Mask-RCNN }    & ${0.765} \, (0.292)$ & $0.093 \,  (0.166)$ & $0.037 \,  (0.112)$ \\ \hline
\textsc{SAM }          & ${0.547} \, (0.186)$ & $0.282 \,  (0.112)$ & $0.016 \,  (0.088)$ \\ \hline
\end{tabular}\label{tab:all_comparison_K3_DSC}
\end{table}


\subsection{Experiments using real data}

This section studies the effectiveness of the proposed approach for real data {segmentation}. The proposed segmentation method is {first} applied to a mammogram {image} for tumor segmentation in breast screening\RCH{, and second, to a synthetic aperture radar (SAR) image for water body detection}.

Mammography remains the primary technique for detecting and diagnosing breast cancer. Early cancer detection can significantly reduce treatment costs. Therefore, the development of methodologies that enable accurate tumor detection is of great interest, with promising results obtained using multifractal-based methods \cite{lopes2009fractal}. \RCH{On the other hand, water body segmentation based on SAR images has become a research hotspot in remote sensing applications given that SAR satellites are able to operate any time, regardless of weather or lighting conditions
\cite{kim2021large,chen2023towards}. However, SAR images are contaminated by multiplicative speckle noise, making their use difficult for computer vision tasks \cite{chen2023towards}.}
\RCH{From these scenarios, it is important to remark that the applicability of the proposed method can be determined by assessing whether the data exhibit multifractal, scale-free behavior. This can be verified by checking if the scaling relation \eqref{eq_cumulant} \cite{combrexelle2015bayesian} holds empirically, which is the case for the selected scenarios.
} 
\begin{figure}[t]
    \centering    
    \includegraphics[width=1\linewidth]{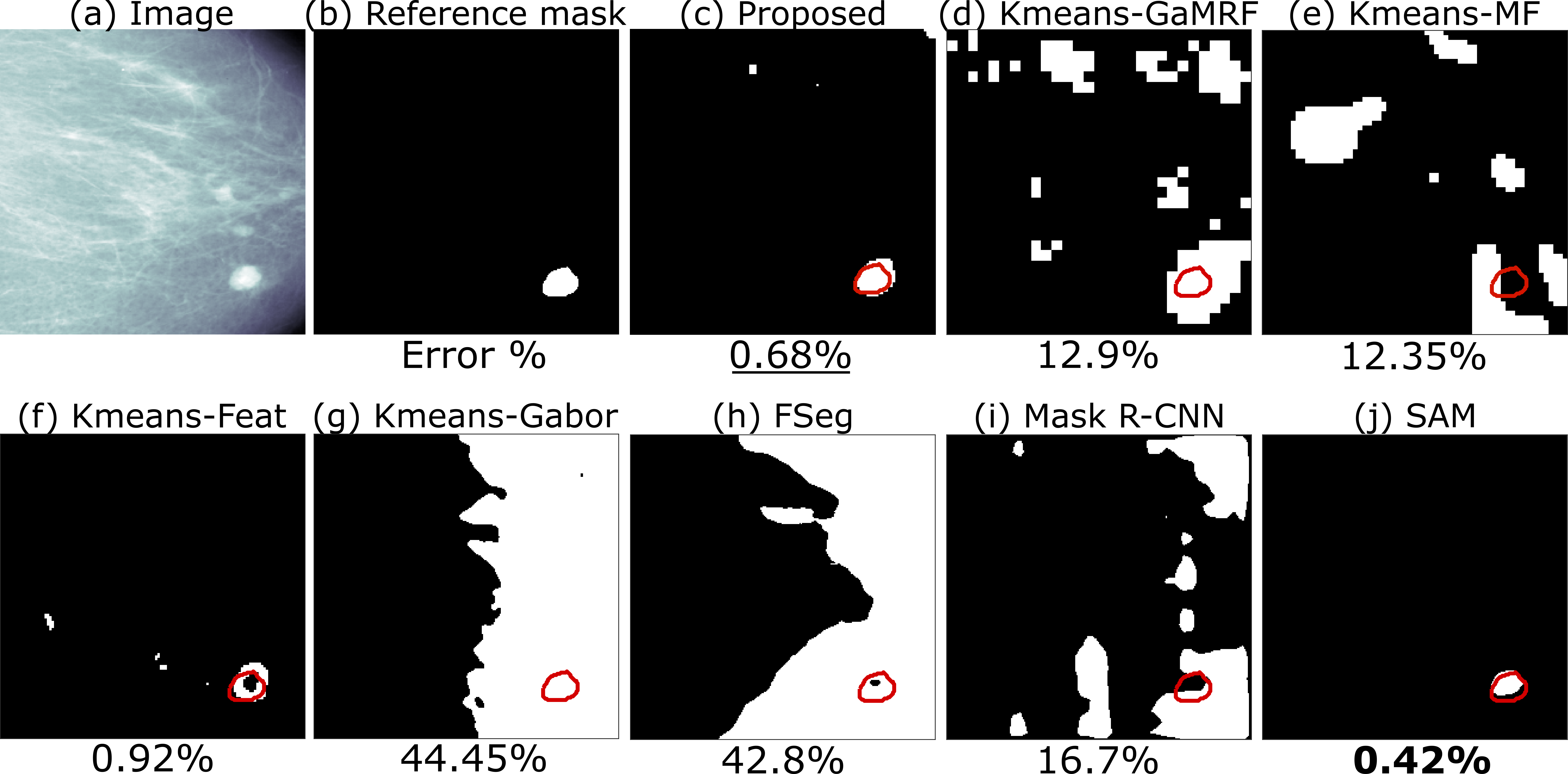} 
    \caption{Comparison of segmentation results on a mammography image for tumor detection, with the tumor contour shown in red on the resulting masks and the segmentation error displayed below each result. Notably, SAM method achieves the best segmentation performance, yet it requires ground-truth pixels from each region as prompts for its operation, making it supervised. The second best segmentation is achieved with the proposed method, which is fully unsupervised, showing its capability of modeling different multifractalities on an image for segmentation purposes.}  
    \label{fig:image_mask}
\end{figure}

\subsubsection{Tumor segmentation}

The mammography image used for this experiment was {extracted} from the Mini-DDSM (Digital Database of Screening Mammography) dataset\footnote{Mammography dataset available at \url{https://ardisdataset.github.io/MiniDDSM/}. The set used for the simulation is ``cancer/1727".}. In this dataset, each image has a segmentation label, or contour, that delineates the margins of the tumor\footnote{It is important to note that the provided contour serves just as a reference to localize the tumor, as literature has reported that such contours may exhibit inaccuracies. Nevertheless, they offer a sufficiently accurate approximation for characterizing the tumor. The segmentation mask is computed from the provided contour by performing a flood-fill operation on the binary mask.}. \RCH{In this context, the proposed method {can be considered with $K=2$ assuming that tumor and normal tissue are characterized by different multifractal parameters}.} The mammography image {was} cropped to retain only the region of interest shown in Figure \ref{fig:image_mask}-(a), removing the surrounding background. Figure \ref{fig:image_mask} shows the analyzed section of (a) the mammography image with (b) its respective segmentation mask, highlighting the tumor, and (c-j) the segmentation results of the proposed method compared to the baseline methods. The proposed method yields very satisfactory results, with the second best error (0.68\%) after the SAM method (0.42\%). Note that SAM produces more false negatives in the region of interest (\textsc{FN:} $203$) than the proposed method (\textsc{FN:} $53$).  Also note that SAM operates in a supervised manner and requires knowledge of input pixels from each region. In contrast, the proposed method operates in a fully unsupervised mode, demonstrating its potential effectiveness in modeling and handling tumor segmentation tasks.

\subsubsection{\RCH{Water detection from SAR images}} 
\begin{figure*}[t]
    \centering
    \includegraphics[width=1\linewidth]{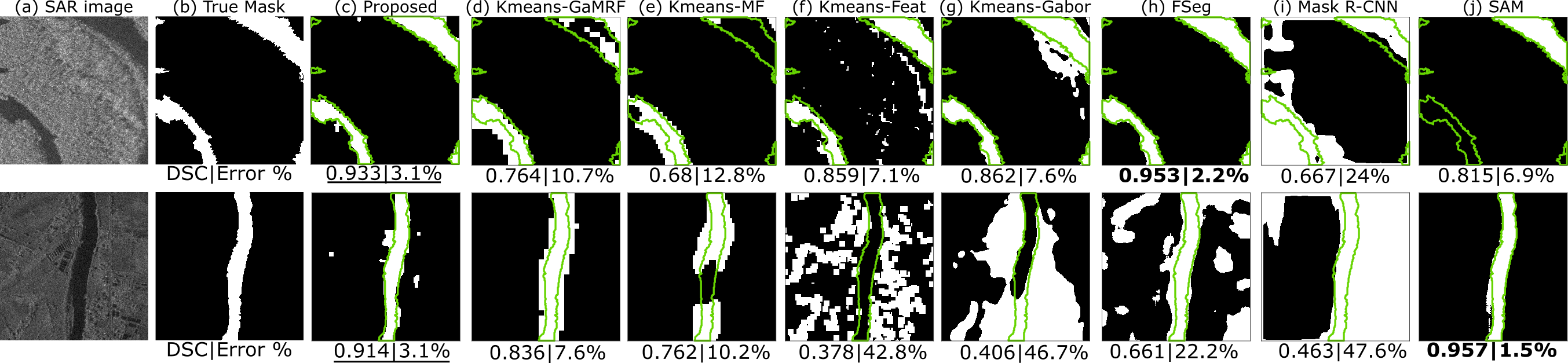} 
    \caption{\RCH{Results on SAR images for segmenting water bodies. First and second best performances are highlighted in bold and with an underline, respectively. For SAM, two points per class were provided as prompts to guide the segmentation. All the other methods are unsupervised.}}
    \label{fig:SAR_results}
\end{figure*}

\RCH{Two SAR images were extracted from the dataset in \cite{kim2021large}, where the true label for the water segmentation is provided (we considered $K=2$). Figure \ref{fig:SAR_results} presents a comparison of segmentation performance, showing that the proposed unsupervised method (c) achieves competitive results and is second best after the \textsc{FSeg} (h) and \textsc{SAM} methods (j): \textsc{SAM}, which is supervised (two points per class were provided as input), yields better performance for the second image but significantly worse for the first, where it fails to detect one of the two water bodies. The \textsc{FSeg} method shows outstanding performance for the first image, but fails quite drastically for the second one. In contrast, the proposed method well detects the water bodies in all cases and yields very satisfactory and robust performance. These results underline the robustness and versatility of the proposed method to handle different segmentation scenarios on challenging images such as SAR data.
}

\section{Conclusions}

This work introduced a new  Bayesian multifractal image segmentation method formulated as the joint estimation of the multifractality parameters and pixel labels. 
As a novel key ingredient, the proposed approach leverages on a flexible Fourier domain model for log-leaders allowing the multifractality estimation of heterogeneous images. To model the space-scale label correlations, a multiscale graph structure was introduced and used in a Potts Markov random field prior. A Bayesian formulation was finally introduced to estimate the unknown model parameters (including the multifractality of each region of the image, the image labels and the granularity parameters of the image regions), numerically and efficiently computed using samples generated by an appropriate Gibbs sampler. 
Numerical experiments conducted on synthetic multifractal images and real data demonstrate the excellent performance of the proposed approach for image segmentation, outperforming several baselines including local multifractality-based approaches, classical texture segmentation methods as well as recent deep learning strategies. 
Future work includes \RCHH{reducing computational time through parallel implementations (e.g., GPUs) and} the extension of the model to data with more than two dimensions (such as depth or spectral data volumes \cite{belmekki20233d}), in which multidimensional/multivariate correlations can be included and exploited in the model. \RCHH{Other} potential directions \RCHH{are the evaluation of} the proposed approach in scenarios with corrupted spatial information, such as noisy, missing or compressed data (as in \cite{leon2020online})\RCHH{, and the integration of the proposed method with deep learning networks (as in \cite{xiong2025fractal})}.

\bibliographystyle{IEEEtran}
\bibliography{biblio}

\end{document}